\documentclass[aps,pr,tightenlines,superscriptaddress,showpacs,amsmath,amssymb]{revtex4}
\usepackage{graphicx} % Include figure files
\usepackage{color} % color
\usepackage{dcolumn}  % Align table columns on decimal point
\usepackage{epsfig}
\usepackage{refmerge}

\def \lr #1{\left( #1 \right)}

\def \B {{\cal B}}

\def \GeV {{\rm GeV}}
\def \MeV {{\rm MeV}}

\def \simlt {\stackrel{<}{\sim}}
\def \simgt {\stackrel{>}{\sim}}

\begin{document}

\vspace*{-1.2\baselineskip}
%\resizebox{!}{3cm}{\includegraphics{belle.eps}}

\includegraphics[width=3.5cm]{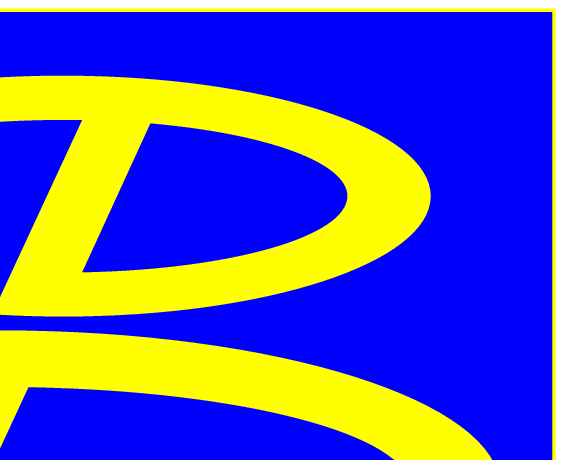}\\
\vspace*{-3.0cm}
\ \\
\hspace*{13.5cm} {\bf Belle Preprint 2008-12}\\
\hspace*{13.5cm} {\bf KEK Preprint 2008-6}\\
\hspace*{15.7cm} {\bf May 2008}\\
\hspace*{14.1cm} {\bf (Revised\ Sep\ 2008) }\\
\ \\
\vspace{0.5cm}
%\preprint{\vbox{ 
%\hbox{   }
%\hbox{  }
%%%%%                  \hbox{BELLE-PUB-DRAFT      }
%                 \hbox{LP2005-xx}
%                 \hbox{EPS05-yy} 
%                 \hbox{hep-ex nnnn, if available}
%%%%%                 \hbox{Contact: S.Uehara}
%%%%%                  \hbox{V3.4--6-Dec-2005}
%%%%%                  \hbox{BELLE-PUB-DRAFT      }
%%                  \hbox{\large Belle Preprint 2005-**}
%%                 \hbox{\large KEK Preprint 2005-** }
%%                  \hbox{\large BELLE-CONF-0661}
%                  \hbox{\large Belle Preprint 2006-xx}
%                 \hbox{\large KEK Preprint 2006-xx }
%                 \hbox{October 6, 2006}
%                 \hbox{{\small *DRAFT*(v3.70)}}
%}}
%
%\vspace*{-1.2cm}
\pagestyle{plain}
\title{ 
%\quad\\[0.5cm]  \ \\ \ \\
High-statistics measurement of neutral-pion pair production in 
two-photon collisions\\
}

%{\color{red} xxxx}
\begin{abstract}
We present a high-statistics measurement of 
differential cross sections
and the total cross section
for the process $\gamma \gamma \to \pi^0 \pi^0$ in the kinematic range
 0.6~GeV $ \leq W \leq 4.0$~GeV and $|\cos \theta^*| \leq 0.8$,
where $W$ and $\theta^*$ are the energy and pion scattering angle, 
respectively, in the $\gamma\gamma$ center-of-mass system. 
Differential cross sections are fitted to obtain information on 
S, D$_0$, D$_2$, G$_0$ and G$_2$ waves.
The G waves are important above $W \simeq 1.6~\GeV$.
General behavior of partial waves is studied by fitting differential
cross sections in a simple parameterization where amplitudes contain
resonant contributions and smooth background. The D$_2$ wave is dominated by the
$f_2(1270)$ meson whose parameters are consistent with the with the current 
world averages. The D$_0$ wave contains a $f_2(1270)$ component, whose fraction 
is fitted. For the S wave, the $f_0(980)$ parameters are found to be consistent
with the values determined from our recent $\pi^+ \pi^-$ data. In addition to 
the $f_0(980)$, the S wave prefers to have another resonance-like contribution 
whose parameters are obtained.

%For $W \leq 1.6~\GeV$ the D$_2$ wave is dominated by
%the $f_2\lr{1270}$ resonance while the S wave requires at least one 
%additional resonance besides the $f_0\lr{980}$, which may be the 
%$f_0(1370)$ or $f_0(1500)$.
%The differential cross sections are fitted with a simple parameterization
%to determine the parameters (the mass, total width and 
%$\Gamma_{\gamma \gamma}{\cal B}(f_0 \rightarrow \pi^0 \pi^0)$)
%of this scalar meson as well as the $f_0(980)$.
%The helicity 0 fraction of the $f_2(1270)$ meson,
%taking into account interference for the first time, is also obtained.

%The mass, total width and $\Gamma_{\gamma \gamma}{\cal B}(f_0 \rightarrow
%\pi^0 \pi^0)$ are obtained to be $(1237~^{+4}_{-3}~^{+243}_{-30})~\MeV/c^2$,
%$(184 \pm 4~^{+135}_{-66})$~MeV and $(467~^{+91}_{-60}~^{+2392}_{-339})$~eV, 
%respectively.
%Also the mass, $\Gamma_{\gamma \gamma}$  and $\Gamma_{\pi \pi}$ of the
%$f_0(980) $ are fitted to be 
%$(977.4~^{+1.4}_{-1.0}~^{+10.7}_{-4.2})~\MeV/c^2$,
%$(314~^{+17}_{-19}~^{+241}_{-199})$~eV and 
%$(67.3 \pm 2.1 ~^{+25.0}_{-31.0})$~eV, respectively.
\end{abstract}

\affiliation{Budker Institute of Nuclear Physics, Novosibirsk}
\affiliation{Chiba University, Chiba}
\affiliation{University of Cincinnati, Cincinnati, Ohio 45221}
\affiliation{Department of Physics, Fu Jen Catholic University, Taipei}
\affiliation{Justus-Liebig-Universit\"at Gie\ss{}en, Gie\ss{}en}
\affiliation{The Graduate University for Advanced Studies, Hayama}
%%%\affiliation{Gyeongsang National University, Chinju}
\affiliation{Hanyang University, Seoul}
\affiliation{University of Hawaii, Honolulu, Hawaii 96822}
\affiliation{High Energy Accelerator Research Organization (KEK), Tsukuba}
\affiliation{Hiroshima Institute of Technology, Hiroshima}
%%%\affiliation{University of Illinois at Urbana-Champaign, Urbana, Illinois 61801}
\affiliation{Institute of High Energy Physics, Chinese Academy of Sciences, Beijing}
\affiliation{Institute of High Energy Physics, Vienna}
\affiliation{Institute of High Energy Physics, Protvino}
\affiliation{Institute for Theoretical and Experimental Physics, Moscow}
\affiliation{J. Stefan Institute, Ljubljana}
\affiliation{Kanagawa University, Yokohama}
\affiliation{Korea University, Seoul}
%%%\affiliation{Kyoto University, Kyoto}
\affiliation{Kyungpook National University, Taegu}
\affiliation{\'Ecole Polytechnique F\'ed\'erale de Lausanne (EPFL), Lausanne}
\affiliation{Faculty of Mathematics and Physics, University of Ljubljana, Ljubljana}
\affiliation{University of Maribor, Maribor}
\affiliation{University of Melbourne, School of Physics, Victoria 3010}
\affiliation{Nagoya University, Nagoya}
\affiliation{Nara Women's University, Nara}
\affiliation{National Central University, Chung-li}
\affiliation{National United University, Miao Li}
\affiliation{Department of Physics, National Taiwan University, Taipei}
\affiliation{H. Niewodniczanski Institute of Nuclear Physics, Krakow}
\affiliation{Nippon Dental University, Niigata}
\affiliation{Niigata University, Niigata}
\affiliation{University of Nova Gorica, Nova Gorica}
\affiliation{Osaka City University, Osaka}
%%%\affiliation{Osaka University, Osaka}
\affiliation{Panjab University, Chandigarh}
%%%\affiliation{Peking University, Beijing}
%%%\affiliation{Princeton University, Princeton, New Jersey 08544}
%%%\affiliation{RIKEN BNL Research Center, Upton, New York 11973}
\affiliation{Saga University, Saga}
\affiliation{University of Science and Technology of China, Hefei}
\affiliation{Seoul National University, Seoul}
%%%\affiliation{Shinshu University, Nagano}
\affiliation{Sungkyunkwan University, Suwon}
\affiliation{University of Sydney, Sydney, New South Wales}
%%%\affiliation{Tata Institute of Fundamental Research, Mumbai}
\affiliation{Toho University, Funabashi}
\affiliation{Tohoku Gakuin University, Tagajo}
%%%\affiliation{Tohoku University, Sendai}
\affiliation{Department of Physics, University of Tokyo, Tokyo}
%%%\affiliation{Tokyo Institute of Technology, Tokyo}
\affiliation{Tokyo Metropolitan University, Tokyo}
\affiliation{Tokyo University of Agriculture and Technology, Tokyo}
%%%\affiliation{Toyama National College of Maritime Technology, Toyama}
\affiliation{Virginia Polytechnic Institute and State University, Blacksburg, Virginia 24061}
\affiliation{Yonsei University, Seoul}

\author{S.~Uehara}\affiliation{High Energy Accelerator Research Organization (KEK), Tsukuba} % KEK
\author{Y.~Watanabe}\affiliation{Kanagawa University, Yokohama} % Kanagawa
\author{I.~Adachi}\affiliation{High Energy Accelerator Research Organization (KEK), Tsukuba} % KEK
\author{H.~Aihara}\affiliation{Department of Physics, University of Tokyo, Tokyo} % Tokyo
%\author{D.~Anipko}\affiliation{Budker Institute of Nuclear Physics, Novosibirsk} %BINP
\author{K.~Arinstein}\affiliation{Budker Institute of Nuclear Physics, Novosibirsk} % BINP
%\author{T.~Aso}\affiliation{Toyama National College of Maritime Technology, Toyama} % Toyama
%\author{V.~Aulchenko}\affiliation{Budker Institute of Nuclear Physics, Novosibirsk} % BINP
%\author{T.~Aushev}\affiliation{\'Ecole Polytechnique F\'ed\'erale de Lausanne (EPFL), Lausanne}
%%\affiliation{Institute for Theoretical and Experimental Physics,Moscow} % ITEP
%\author{T.~Aziz}\affiliation{Tata Institute of Fundamental Research, Mumbai} % Tata
%\author{S.~Bahinipati}\affiliation{University of Cincinnati, Cincinnati, Ohio 45221} % Cincinnati
\author{A.~M.~Bakich}\affiliation{University of Sydney, Sydney, New South Wales} %Sydney
\author{V.~Balagura}\affiliation{Institute for Theoretical and Experimental Physics, Moscow} % ITEP
%\author{Y.~Ban}\affiliation{Peking University, Beijing} % Peking
\author{E.~Barberio}\affiliation{University of Melbourne, School of Physics, Victoria 3010} % Melbourne
%\author{M.~Barbero}\affiliation{University of Hawaii, Honolulu, Hawaii 96822} %Hawaii
\author{A.~Bay}\affiliation{\'Ecole Polytechnique F\'ed\'erale de Lausanne (EPFL), Lausanne} % Lausanne
\author{I.~Bedny}\affiliation{Budker Institute of Nuclear Physics, Novosibirsk} %BINP
% ...
% BINP
\collaboration{The Belle Collaboration}

%\vspace*{\baselineskip}
\medskip
%\date{April 16, 2008, Ver.~5.0}

%\noaffiliation

\maketitle
%\newpage

%%%% >>>> keep the final version single-spaced
\tighten

%{\renewcommand{\thefootnote}{\fnsymbol{footnote}}}
%\setcounter{footnote}{0}

%\begin{multicols}{2}
\section{Introduction}
\label{sec:intro}
Studies of exclusive hadronic final states in two-photon
collisions give valuable information on the physics of light and heavy-quark
resonances, perturbative and nonperturbative quantum chromodynamics (QCD) 
and hadron-production mechanisms. So far, Belle has measured
the production cross sections of charged-pion pairs
~\cite{bib:mori1,bib:mori2,bib:nkzw}, 
charged- and neutral-kaon pairs~\cite{bib:nkzw,bib:kabe,bib:wtchen}, 
and proton-antiproton pairs~\cite{bib:kuo}.
We have also analyzed $D$-meson-pair production finding a new
charmonium state identified as the $\chi_{c2}(2P)$~\cite{bib:uehara}.

Here we present the cross sections and an analysis of
high-statistics neutral-pion pair production in two-photon processes.
The motivation of this study is essentially the same as that
for charged-pion pair production. 
However, the two processes are physically different and 
independent; we cannot predict one by measuring only the other.

In the low energy region ($W \simlt 0.8$~GeV, where $W$ is the center-of-mass
(c.m.) energy of the $\pi \pi$ system), it is expected 
that the difference of meson electric charges 
plays an essential role in the difference between the $\pi^+\pi^-$ and 
$\pi^0\pi^0$ cross sections. 
Predictions are not straightforward because of non-perturbative effects.
In the intermediate energy range ($0.8~\GeV \simlt W \simlt 2.4$~GeV), 
formation of meson resonances decaying to $\pi\pi$ is the dominant 
contribution.  
Since ordinary $q\bar{q}$ mesons conserve isospin in 
decays to $\pi\pi$, we can restrict  the $I^G J^{PC}$ quantum numbers
of the meson produced by two photons to be  
$0^+$(even)$^{++}$, that is, $f_{J={\rm even}}$ mesons. 
The ratio of the $f$-meson's branching fractions,
${\B}(f \to \pi^0\pi^0)/{\B}(f \to \pi^+\pi^-)$
is 1/2 from isospin invariance. 
However, interference of resonances 
with the continuum component, which cannot be precisely 
calculated, distorts this ratio even near the resonant peaks.

A long-standing puzzle in QCD
is the existence and structure of low mass 
($\simlt 1.5~\GeV/c^2$) scalar mesons~\cite{bib:scalar}.
They, in particular, the $f_0(600)$ ($\sigma$ meson), are closely related
to the QCD vacuum through spontaneous breakdown of chiral symmetry.
Two-photon-resonant production of a meson gives valuable information such as
its two-photon width, which is sensitive to its charge structure.
Unfortunately, experimental constraints in a $B$ factory experiment
do not allow measurements much below $W \simlt 0.8$~GeV.

For higher energies ($W \simgt 2.4~\GeV$), we can invoke a quark model.
In leading order calculations~\cite{bib:bl,bib:bc}
which take into account spin correlations between quarks, the $\pi^0\pi^0$
cross section is predicted to be much smaller
than that of $\pi^+\pi^-$, and the ratio 
of  $\pi^0\pi^0$ to $\pi^+\pi^-$ is
around 0.03-0.06. 
However, higher-order or nonperturbative QCD effects can modify this ratio. 
For example, the handbag model, which considers soft hadron exchange, predicts
the same amplitude for the two processes
and hence this ratio is expected to be 0.5~\cite{bib:handbag}. 
Analyses of energy and angular distributions of the cross sections
are essential to determine properties of the observed
resonances and to test the validity of QCD models.

In this paper, we present measurements of the differential cross sections, 
$d\sigma/d|\cos \theta^*|$,
for the process $\gamma \gamma \to \pi^0 \pi^0$ in
a wide two-photon c.m. energy ($W$) range from 0.6 to 4.0~GeV,
in the c.m. angular range, $|\cos \theta^*| \leq 0.8$.
%(Although we do not need to put the absolute-value symbol
%for $\cos \theta^*$ for the present channel where the identical
%particle pairs appear in both initial and final states,
%we follow the usual convention of the two-photon differential
%cross section to avoid unnecessary confusion.)
The 95~fb$^{-1}$ data sample results in several hundred times larger
statistics than in previous experiments~\cite{bib:prev1, bib:prev2}.
The data here are concentrated in the low energy region
($W \leq 1.6~\GeV$).
In the higher energy region, many more resonances contribute and thus make
an analysis based on a single experiment difficult.
Furthermore, statistics are still poor in this range.
The angular coverage up to $|\cos \theta^*| = 0.8$ greatly enhances the
capability for separating partial waves.
In the low energy region, the cross section is dominated by the $f_2(1270)$.
A clear peak corresponding to the $f_0(980)$ is found for the first time
in two-photon production of $\pi^0 \pi^0$.
Furthermore, a ``model-independent'' partial wave analysis (where
interference terms of amplitudes are temporarily neglected) reveals another
resonance-like structure around 1.2~GeV in the S wave.
This may be due to the contributions of the $f_2(1270)$ in the D$_0$ wave 
and/or scalar resonances such as the $f_0(1370)$.
We fit the differential cross sections assuming such contributions and obtain
their parameters.

This paper is organized as follows.
In section~\ref{sec:apparatus}, the experimental apparatus relevant to this
measurement is briefly described together with information on
the trigger and a description of the event selection.
Differential cross sections are derived in section~\ref{sec:crosssec}.
In section ~\ref{sec:fitting}, the measured differential cross sections are 
fitted to obtain the resonance parameters of the $f_0(980)$ and a scalar 
meson as well as to extract the fraction of the $f_2(1270)$ in the D$_0$ wave.
Finally in section~\ref{sec:summary} a summary and conclusion are given.
%Appendix~\ref{sec:appendix} gives the partial wave amplitudes
%when interference effects are taken into account.

\section{Experimental Apparatus and Event Selection}
\label{sec:apparatus}
%Here the experimental apparatus and trigger system relevant to the 
%measurement are briefly described.
%Then an account on the event selection is given.

%\subsection{Data Sample}
We use data that corresponds to an integrated luminosity of
95~fb$^{-1}$ recorded with the Belle detector at the KEKB 
asymmetric-energy $e^+e^-$ collider~\cite{bib:kekb}. 
The $e^+e^-$ c.m. energy 
of the accelerator was set at 10.58~GeV (83~fb$^{-1}$),
10.52~GeV (9~fb$^{-1}$), 10.36~GeV ($\Upsilon(3S)$ runs,
2.9~fb$^{-1}$) and 10.30~GeV (0.3~fb$^{-1}$).
The differences between the two-photon flux (luminosity function) 
in the measured $W$ regions due to differences in
the beam energies are small (at most a few percent), and
the fraction of integrated luminosity of the runs with
lower beam energies is also small; we combine the results for
different beam energies. 
The variation of the cross section because of this effect
is less than 0.5\%.

%\subsection{Experimental Apparatus}
The Belle detector is a magnetic spectrometer covering a large
solid angle (polar angles between $17^{\circ}$ and $150^{\circ}$ and 
the full azimuthal angle).
% with high momentum resolution and good particle identification capability.
A comprehensive description of the Belle detector is
given elsewhere~\cite{bib:belle}. We mention here only those
detector components that are essential to the present measurement.
Charged tracks are reconstructed from hit information in a central
drift chamber (CDC) located in a uniform 1.5~T solenoidal magnetic field.
The $z$ axis of the detector and the solenoid are along the positron beam
direction, with the positrons moving in the $-z$ direction.  
The CDC measures the longitudinal and transverse-momentum components 
(along the $z$ axis and in the $r\varphi$ plane, respectively).  
Photons are detected and measured in an
electromagnetic calorimeter (ECL) located inside the solenoid.
The ECL is an array of 8736 CsI(Tl) crystals pointing toward
the interaction point, which help separating photons from $\pi^0$'s
for energies up to $\simeq 4~\GeV$.

%\subsection{Trigger}
We require that there be no
reconstructed CDC tracks coming from the vicinity of
the nominal collision point. 
Photons from decays of two neutral pions are measured in the ECL. 
Signals from the ECL are used to trigger.
The ECL trigger requirements
are the following: the total ECL energy deposit
in the triggerable acceptance region (see below)
is greater than 1.15~GeV (the ``HiE'' trigger) or
the number of ECL clusters (each crystal having more than 110~MeV)
is four or greater (the ``Clst4'' trigger).
The above energy thresholds are determined from
a study of the correlations between the two triggers in data.  
This trigger logic is realized in hardware;
no additional software filters are applied for events triggered by either
of the two ECL triggers.

%\subsection{Event Selection}
The analysis is performed in the ``zero-tag'' mode, where
neither the recoil electron nor the positron is detected. We
restrict the virtuality of the incident photons to be small
by imposing 
strict transverse-momentum balance with respect to the beam axis
for the final-state hadronic system.

The selection conditions for $\gamma \gamma \to \pi^0 \pi^0$ 
signal candidates are the following.
All the variables in criteria (1)-(6) are measured in
the laboratory frame: (1) there is no good track 
that satisfies $dr < 5$~cm, $dz < 5$~cm and $p_t > 0.1$~GeV/$c$,
where $dr$ and $dz$ are the radial and axial distances,
respectively, of closest approach (as seen in the $r\varphi$ plane) 
to the nominal collision point, and  the $p_t$ is the transverse 
momentum measured in the laboratory frame with respect to the $z$ axis; 
%%All the variables 
%%in criteria (1)-(5) are measured in the laboratory frame;
(2) the events are triggered by either the HiE or Clst4 triggers;
(3) there are two or more photons whose energies are greater than 100~MeV;
(4) there are exactly two $\pi^0$'s, each $\pi^0$ having a transverse momentum 
greater than 0.15~GeV/$c$
with each of the decay-product photons having an energy greater than 70~MeV;
(5) the two photons' momenta are 
recalculated using a $\pi^0$-mass-constrained fit,
and required to have a minimum $\chi^2$ value for the fit
(there was a negligible fraction of events with ambiguous photon 
combinations); 
(6) the total energy deposit in the ECL is smaller than 5.7~GeV.

The transverse momentum in the $e^+e^-$ c.m. frame
($|\Sigma \mbox{\boldmath$p$}_t^*|$) of the two-pion system
is then calculated. 
For further analysis, 
(7) we use events with $|\Sigma \mbox{\boldmath$p$}_t^*| < 50$~MeV/$c$
as the signal candidates. 

In order to reduce
uncertainty from the efficiency of the hardware ECL triggers, 
we set offline selection criteria that emulate the hardware trigger 
conditions as follows: 
(8) the ECL energy sum within the triggerable region
is greater than 1.25~GeV, {\it or}  all four photons
composing the two $\pi^0$ are contained in
the triggerable acceptance region. Here, we
define the triggerable acceptance region as 
the polar-angle ($\theta$) range in the laboratory system 
$17.0^\circ < \theta < 128.7^\circ$.

\section{Derivation of Differential Cross Sections}
\label{sec:crosssec}
In this section, we describe the derivation of differential
cross sections.
First, candidate events are divided into bins of 
$W$ and $| \cos \theta^*|$.
Backgrounds are then subtracted by fitting the transverse-momentum distribution.
Event distributions are unfolded to correct for finite energy resolution.
Finally, differential cross sections are obtained in
bins of $W$ and $|\cos \theta^*|$.

\subsection{Signal Distributions}
We derive the c.m. energy $W$ of the two-photon collision 
from the invariant mass of the two-neutral-pion system.
We calculate the cosine of the scattering angle of 
$\pi^0$ in the $\gamma \gamma$ c.m. frame, $| \cos \theta^*|$.
We then approximate the $e^+e^-$ collision 
axis in the $e^+e^-$ c.m. frame as the reference for this polar angle.
%(we do not know the exact $\gamma\gamma$ collision axis).

\begin{figure}[ht]
\centering
{\epsfig{file=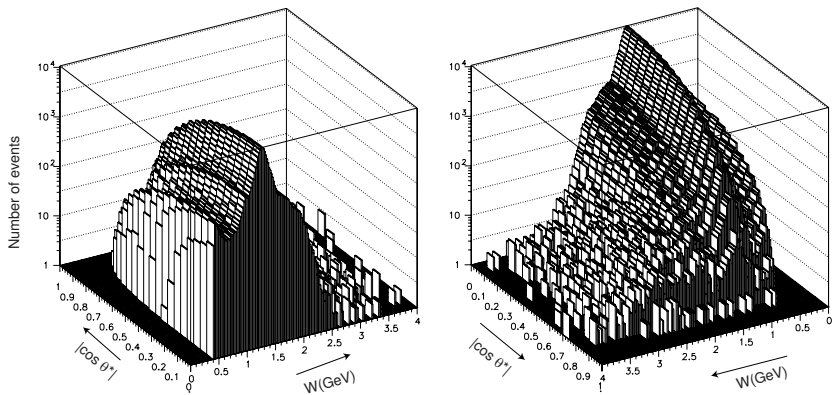, width=11cm}}
\centering
\caption{Two-dimensional $W$ and $|\cos \theta^*|$ distribution
for $\pi^0\pi^0$ candidates in data.
The same distribution is viewed from
two different directions.}
\label{fig:lego}
\end{figure}

\begin{figure}[ht]
\centering
{\epsfig{file=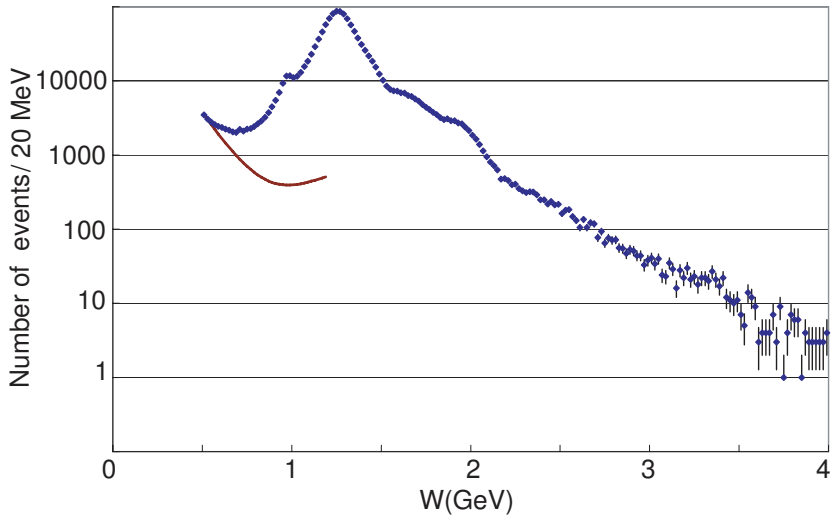, width=11cm}}
\centering
\caption{$W$ distribution
for candidate events. The angular coverage is $|\cos \theta^*| \leq 0.8$.
The curve is an estimate of backgrounds from events with
$p_t$ imbalance (see text).}
\label{fig:wdist}
\end{figure}

The two-dimensional yield distribution of the 
selected events is shown as a lego plot 
in Fig.~\ref{fig:lego}. 
The $W$ distribution with $|\cos \theta^*| \leq 0.8$
is shown in Fig.~\ref{fig:wdist}. 
The total number of events observed is $1.25 \times 10^6$.
We observe clear peaks for the $f_0(980)$ near 0.98~GeV and 
the $f_2(1270)$ near 1.25~GeV and find at least two more structures around 
1.65~GeV and 1.95~GeV. 
%\end{multicols}

\subsection{Background Subtraction}
We use the $p_t$-balance distribution, i.e.,
the distribution in $|\Sigma \mbox{\boldmath$p$}_t^*|$,
to separate the signal and background components.
The signal Monte Carlo (MC) shows that the signal 
component peaks around 10-20~MeV/$c$ in this
distribution.  
In the experimental data, however, in addition to the
signal component, we find some
contributions from $p_t$-unbalanced components 
in the low-$W$ region. 
Such $p_t$-unbalanced backgrounds might originate from processes
such as $\pi^0\pi^0\pi^0$, etc.  
However, the background
found in the experimental data is very large only in the low-$W$
region where the $\pi^0\pi^0\pi^0$ contribution is expected  
to be much smaller than $\pi^0\pi^0$ in two-photon collisions.
(Note that a $C=-$ system cannot decay to $\pi^0\pi^0$.) 
We believe that the backgrounds are dominated by beam-background photons (or
neutral pions from secondary interactions) or spurious hits in the detector.

\begin{figure}[ht]
\centering
{\epsfig{file=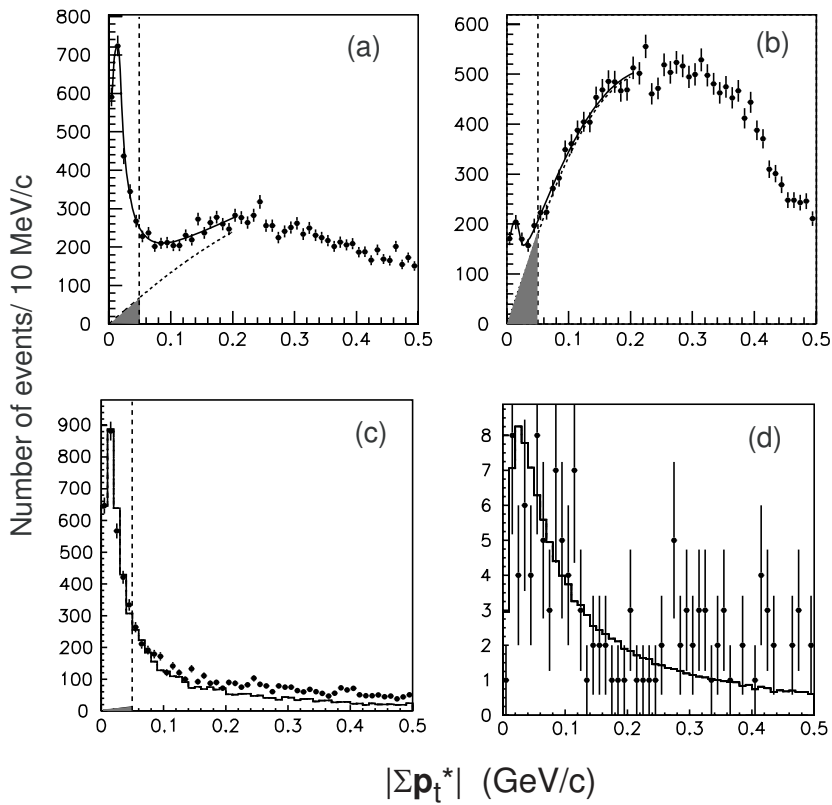, width=11cm}}
\centering
\caption{Distribution of imbalance in $|\sum \mbox{\boldmath$p$}^*_t|$
for candidate events. 
(a) In the bin centered at
$W=0.90~$GeV and $|\cos \theta^*|=0.05$
(the bin width is 0.04~GeV and 0.1 in the $W$ and 
$|\cos \theta^*|$ directions, respectively, in (a)-(c)), 
the experimental distribution (dots with error bars) is fitted 
with the sum of signal and background components (curves). The gray
region shows the estimated background contamination in
the signal region. 
(b) The same as (a) for the bin centered at $W=0.66~$GeV.
(c)  In the $W=1.18$~GeV, $|\cos \theta^*|=0.65$ bin the experimental
distribution is compared with
the signal MC (histogram). 
(d) The same as (c) for $3.6~\GeV \leq W \leq 4.0$~GeV
and  $|\cos \theta^*| \leq 0.4$.}
\label{fig:ptbala}
\end{figure}

Figures \ref{fig:ptbala} (a) and (b) show the $p_t$-balance distributions 
in the low $W$ region. With the fit described below, 
we separate the signal components from the background.
In the intermediate or higher energy regions, the $p_t$ unbalanced
backgrounds
are either less than 1\%, buried under the $f_2(1270)$ peak 
(Fig.~\ref{fig:ptbala}(c)),
or consistent with zero
within statistical errors.
For the highest energy region 3.6~GeV$\leq W \leq 4.0$~GeV,
we subtract a 3\% background from the yield in each bin 
to account for the background from
the $p_t$-unbalanced components and assign a
systematic error of the same size, although it is not statistically significant
even there (Fig.~\ref{fig:ptbala}(d)).

 A fit to the $p_t$-balance distribution is performed
in the region $|\sum \mbox{\boldmath$p$}_t^*| \leq 0.2~\GeV/c$ 
to separate the signal and background components for 
the $W$ region below 1.2~GeV.  The fit function is a sum
of the signal and background components.
The signal component is an empirical function
reproducing the shape of the signal MC, $y=Ax/(x^{2.1}+B+Cx)$, 
where $x \equiv |\sum \mbox{\boldmath$p$}_t^*|$, $A$, $B$ and $C$ are the
fitting parameters, and $y$ is the distribution.
 This function
has a peak at $x=(\frac{B}{1.1})^{\frac{1}{2.1}}$ and vanishes
at $x=0$ and as $x \to \infty$. 
The shape of the 
background is taken as a linear function $y=ax$ for $x<0.05$~GeV/$c$, which is 
smoothly connected to a quadratic function above $x>0.05$~GeV/$c$.

The background yields obtained from the fits are fitted to a 
smooth two-dimensional function of ($W$, $|\cos \theta^*|$), 
in order not to introduce statistical fluctuations. 
The backgrounds are then subtracted from the experimental 
yield distribution.  
The background yields integrated over
angle are shown in Fig.~\ref{fig:wdist} for $W < 1.2$~GeV.
Above 1.2 GeV, we do not find any statistically significant background 
contributions from the fit.
The correction and systematic errors from background are
summarized in Sect. III.F.
We omit the data points in the small-angle ($|\cos \theta^*| \geq 0.6$) 
region with $W \leq 0.72$~GeV, because there the background
dominates the yield.

\subsection{Unfolding the $W$ Distributions}
We estimate the invariant-mass resolutions from studies 
of the signal MC and data.
We find that the MC events have a relative invariant-mass 
resolution of 1.4\%, which is almost constant over the entire $W$ 
region covered by the present measurement.  
The $\pi^0$ momentum resolution is known to be 
about 15\% worse in the experimental data than in MC 
from a study of the $p_t$-balance distributions.
Moreover, the distribution in the MC is 
asymmetric; it has a longer tail on the lower mass side.
An asymmetric Gaussian function with standard deviations of 
1.9\%$W$ and 1.3\%$W$ on the lower and higher sides of the peak, 
respectively, is used and approximates the smearing reasonably well.  
This invariant-mass resolution is comparable to 
or larger than the $W$ bin width (20~MeV) used 
in Figs.~\ref{fig:lego} and \ref{fig:wdist}. 
We unfold the invariant-mass distribution in each $|\cos \theta^*|$ bin
separately, to correct for migrations of signal 
yields to different $W$ bins in obtaining the true $W$ distribution, 
based on the asymmetric Gaussian smearing described above.
The migration in the $|\cos \theta^*|$  direction is expected to be
small and is neglected.
 
\begin{figure}[ht]
\centering
{\epsfig{file=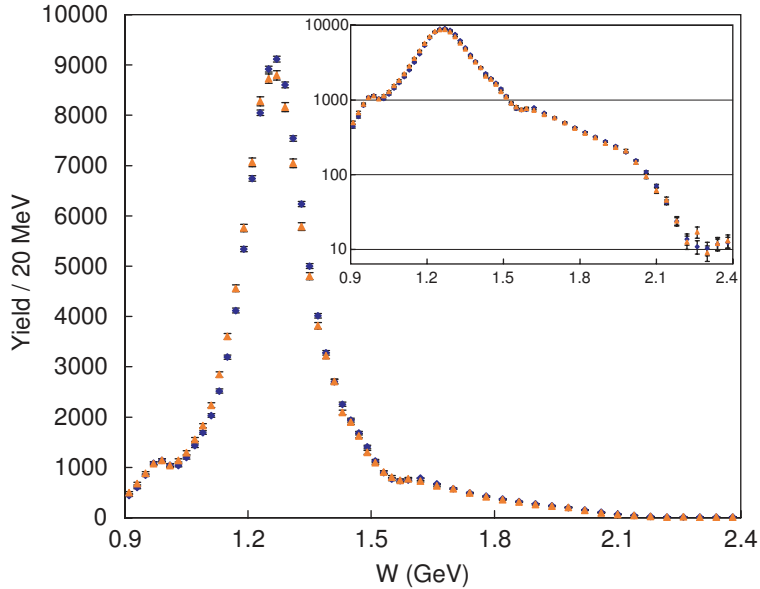, width=10cm}}
\centering
\caption{Invariant-mass distributions
before (orange colored triangles) and
after (dark-blue diamonds) the unfolding,
at $|\cos \theta^*|=0.225$.
The inset shows the same distribution on a semilogarithmic scale. 
The bin width changes from 0.02~GeV
to 0.04~GeV above $W = 1.6$~GeV.}
\label{fig:unfold}
\end{figure}

The unfolding uses the singular value decomposition
algorithm~\cite{bib:svdunf} at the
yield level and is applied so as to 
obtain
the corrected $W$ distribution in the 0.9 - 2.4~GeV region,
using data in the observed $W$ range between 0.72 and 3.0~GeV.
For lower energies, $W \leq 0.9$~GeV, migration is 
expected to be small because of the better
invariant-mass resolution compared
with the bin width. For higher energies, $W > 2.4$~GeV, 
where the statistics are relatively low and
the unfolding would enlarge the errors,
we rebin the data with a bin width of 100~MeV, instead
of unfolding. 
Distributions before and after
the unfolding for a typical angular bin ($|\cos \theta^*|=0.225$) 
are shown in Fig.~\ref{fig:unfold}. 

We calibrate the experimental energy scale and 
invariant-mass distribution using the $\gamma \gamma$
invariant mass from experimental samples 
of $\eta' \to \gamma \gamma$ in two-photon processes.
The peak position is consistent with the nominal mass of
$\eta'$ with an accuracy better than 0.2\%.
The mass resolution estimated from the peak width is also consistent
with the smearing function that is used for the unfolding of the $\pi^0\pi^0$
invariant-mass distributions.

\subsection{Determination of Efficiency}
\begin{figure}[ht]
\centering
{\epsfig{file=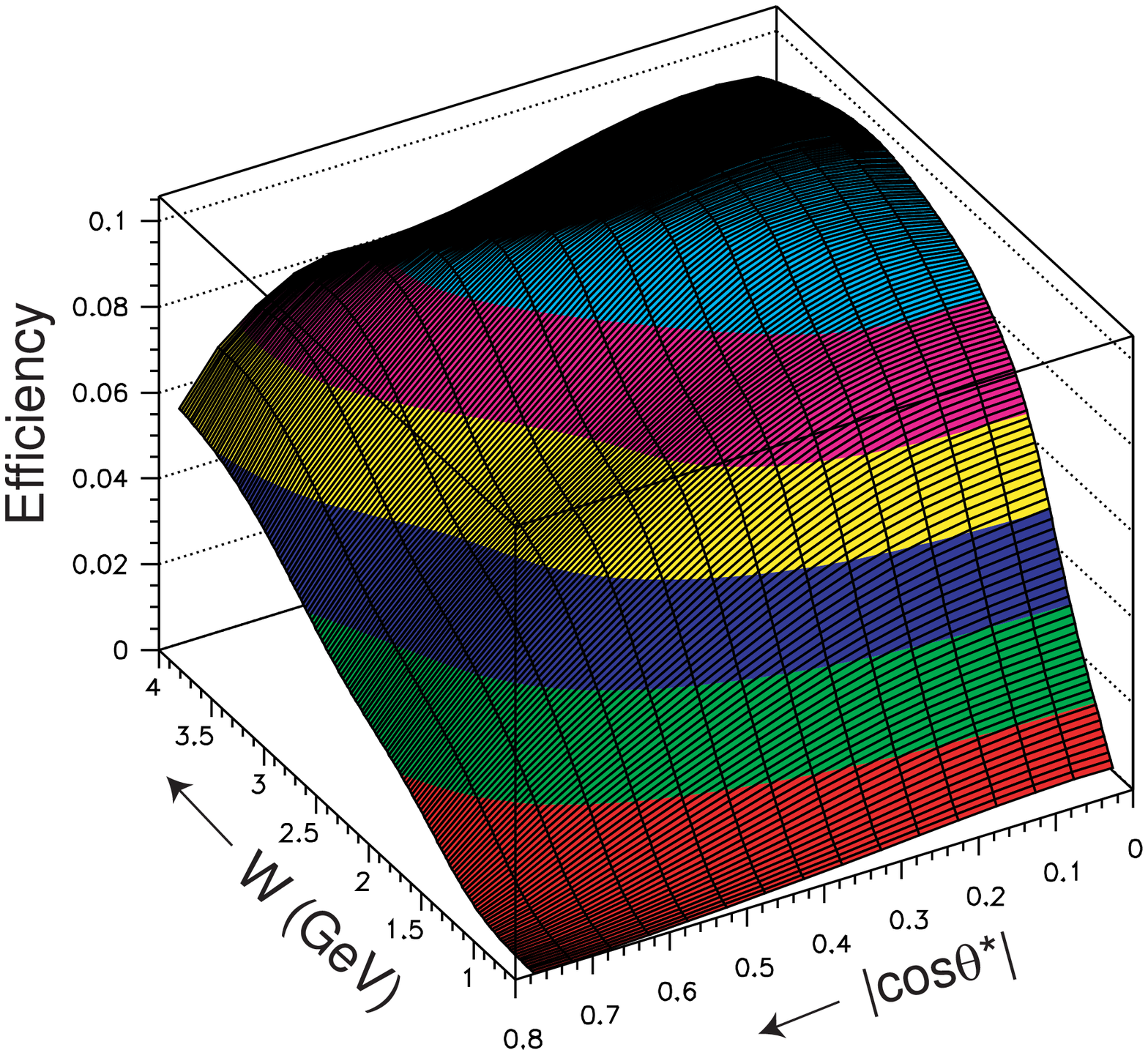, width=11cm}}
\centering
\caption{Overall efficiency in the $W - |\cos \theta^*|$ plane.}
\label{fig:eff08}
\end{figure}

We determine the efficiency for the signal using the detector 
and trigger simulators and applying the selection criteria to signal
MC events.
The $e^+e^- \to e^+e^-\pi^0\pi^0$ signal MC events are generated using
the TREPS code~\cite{bib:treps} at 58 fixed $W$ points between 0.5 and 4.5~GeV,
and isotropic in $|\cos \theta^*|$.  The angular distribution at the
generator level does not play a role in the efficiency determination,
because we calculate the efficiencies separately in each 
0.05 wide $|\cos \theta^*|$ bin.
The number of events generated is $4 \times 10^5$ at each $W$ point.
 To minimize statistical fluctuations in the MC calculation, we fit 
the efficiency to a two-dimensional empirical function in 
($W$, $|\cos \theta^*|$).
The efficiency thus determined is depicted in Fig.~\ref{fig:eff08}. 

\begin{figure}[ht]
\centering
{\epsfig{file=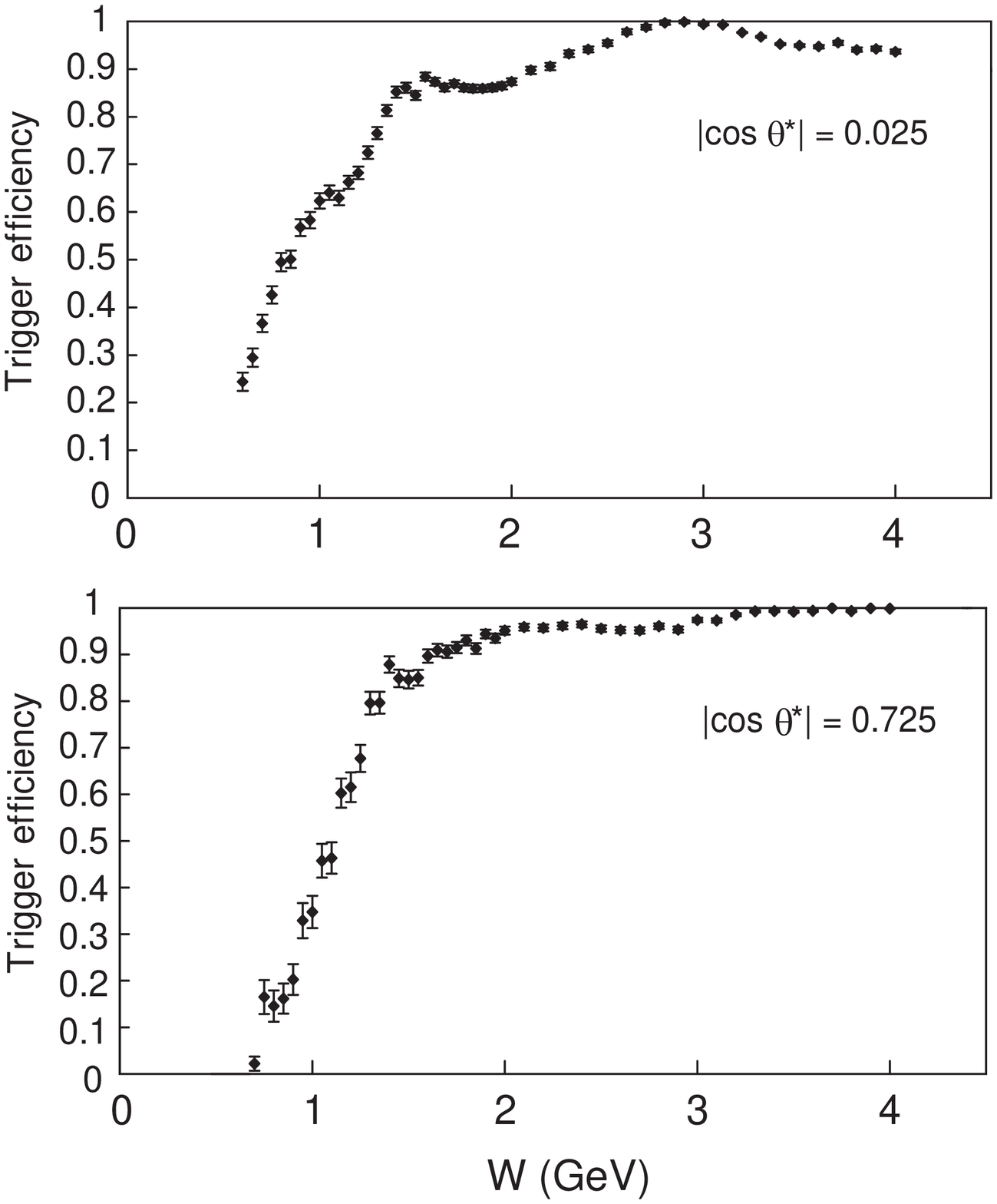, width=11cm}}
\centering
\caption{Trigger efficiencies plotted as a function of $W$ for
two $|\cos \theta^*|$ bins.}
\label{fig:e_trg}
\end{figure}

We find that 
the trigger efficiency, which is defined as the triggerable fraction
of events that can pass through the event selection criteria,
is almost flat and close to 100\% for the region $W \simgt 1.3$~GeV
as shown in Fig.~\ref{fig:e_trg}.
It decreases to typically 30\% at the lowest energies, $W<0.7$~GeV.
Meanwhile, the overall efficiency (the product of the efficiencies 
for the trigger and the acceptance)
is about 11\% at maximum and decreases
(down to around 1\%) at lower $W$ or smaller c.m. angles (larger
$|\cos \theta^*|$).


The overall efficiency calculated using the signal MC events
is corrected for a systematic difference found
between the peak widths in the $p_t$-balance
distributions of the experimental data and
the MC, which could affect the efficiency
through the $|\sum \mbox{\boldmath$p$}_t^*|$ cut. 
This originates from a difference
in the momentum resolution for $\pi^0$'s between data and MC events.
We find that the data peak position
is 10\% to 20\% higher than the MC
expectation, depending on $W$ and $|\cos \theta^*|$. 
The efficiency correction factor ranges from 0.90 to 0.95.

\subsection{Cross-Section Calculation}
The differential cross section for each
($W$, $|\cos \theta^*|$) point is derived
from the following formula:
\begin{equation}
\frac{d\sigma}{d|\cos \theta^*|} =
\frac{\Delta Y - \Delta B}{\Delta W \Delta |\cos \theta^*| 
\int{\cal L}dt L_{\gamma\gamma}(W) \eta }\; ,
\label{eqn:cross}
\end{equation}
where $\Delta Y$ and $\Delta B$ are the signal yield and
the estimated $p_t$-unbalanced background in each bin, 
$\Delta W$ and $\Delta |\cos \theta^*|$ are the bin widths, 
$\int{\cal L}dt$ and  $L_{\gamma\gamma}(W)$ are
the integrated luminosity and two-photon luminosity function
calculated by TREPS, respectively, and  $\eta$
is the net efficiency.
The luminosity function transforms the cross sections for the $e^+e^-$
incident beam to that of the $\gamma\gamma$ incident using a relation:
\[
L_{\gamma\gamma}(W) = \frac{d\sigma_{ee}}{dW}/\sigma_{\gamma\gamma}(W).
\]

The $W$-bin width is 0.02~GeV up to 1.6~GeV and then is modified 
to 0.04~GeV and 0.1~GeV for $W$ between 1.6~GeV and 2.4~GeV and 
$W$ above 2.4~GeV, respectively.
The width of the $|\cos \theta^*|$ bins are 
fixed to $\Delta |\cos \theta^*|=0.05$.

\begin{figure}[ht]
\centering
{\epsfig{file=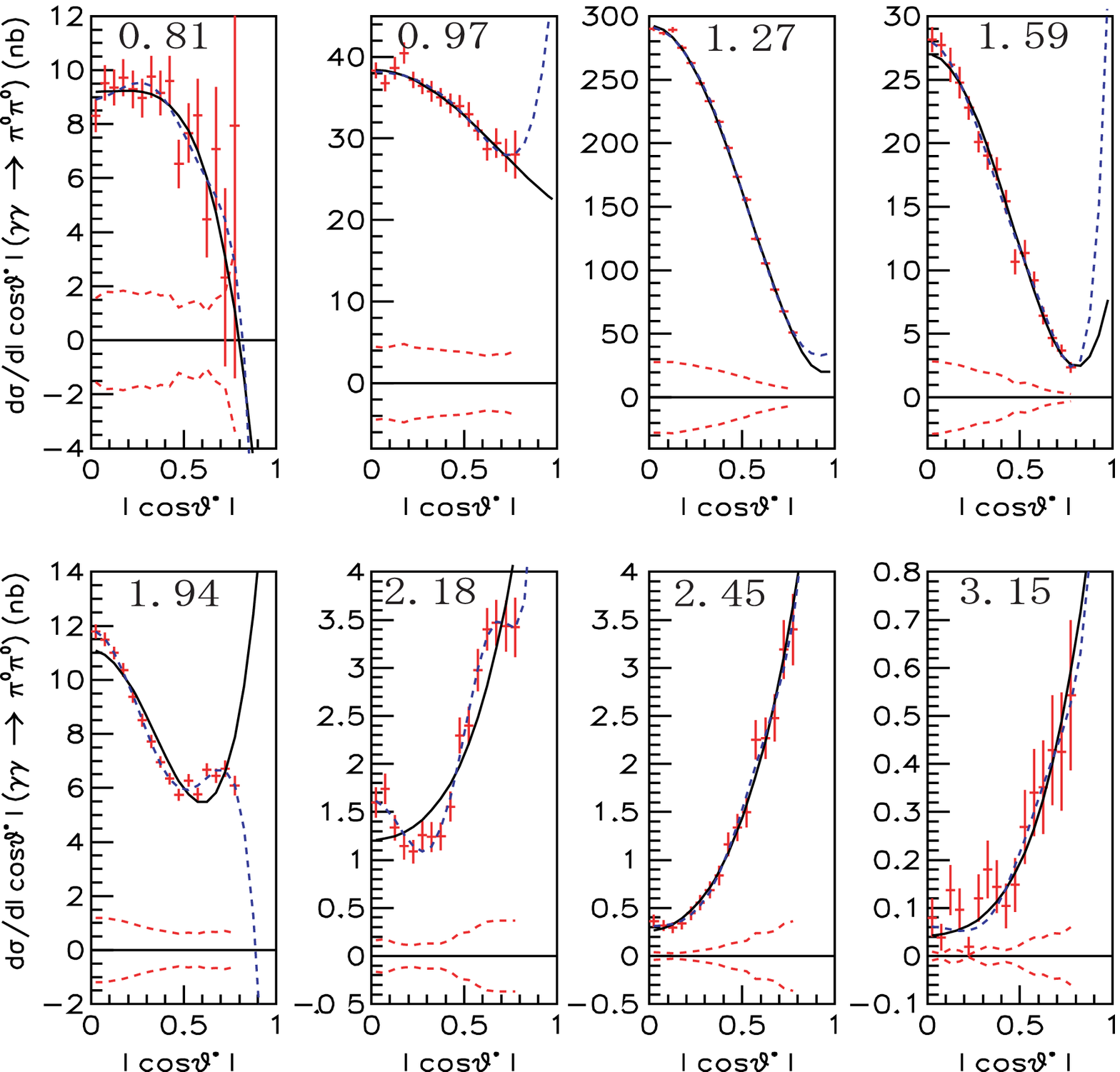, width=11cm}}
\centering
\caption{Differential cross sections for the eight selected $W$ points 
indicated in GeV in each figure.
The results for $W \leq 2.4~\GeV$ are after the unfolding. 
The curves are fit results described in sec.~\ref{sec:fitting};
the solid (dashed) lines are the result of the $SD$ ($SDG$) fit (see text), 
while dotted lines indicate the size of systematic errors in the cross 
sections.}
\label{fig:dssp}
\end{figure}

The differential cross sections obtained are shown
at several $W$ points in Fig.~\ref{fig:dssp}.
They show quite different behaviors.
It should be noted that the cross-section results
after the unfolding are no longer
independent of each other in neighboring bins,
in both central values and sizes of errors.

\begin{figure}[ht]
\centering
{\epsfig{file=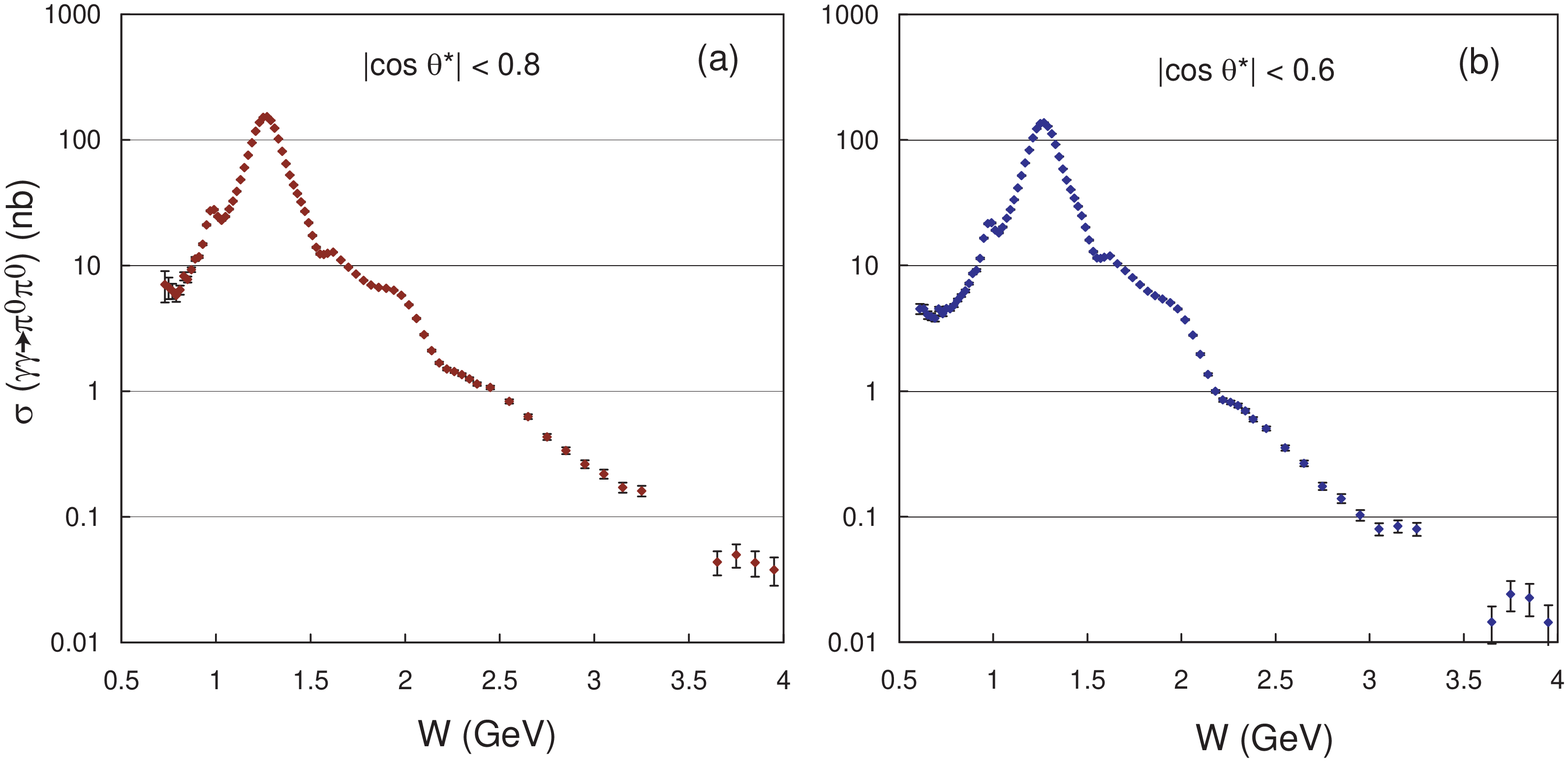, width=11cm}}
\centering
\caption{The total cross section for 
(a) $|\cos \theta^*| \leq 0.8$ and
(b) $|\cos \theta^*| \leq 0.6$ after
the unfolding and rebinning.
See text for the bins in the range 3.3~GeV $\leq W \leq  3.6$~GeV.}
\label{fig:totcs}
\end{figure}

Figures~\ref{fig:totcs}(a) and (b) show the $W$ dependence of the
cross section integrated over $|\cos \theta^*| \leq 0.8$ 
and $|\cos \theta^*| \leq 0.6$, respectively. 
They are obtained by adding
$d\sigma/d|\cos \theta^*| \cdot  \Delta |\cos \theta^*|$ 
over the corresponding angular bins.
The total cross section ($|\cos \theta^*| \leq 0.8$)
is dominated by the $f_2(1270)$.
A clear peak due to the $f_0(980)$ is also visible as in
the $\pi^+ \pi^-$ cross section from Belle~\cite{bib:mori1,bib:mori2}.
Additional structures are visible around $W = 1.65~\GeV$ and $W = 1.95~\GeV$.
The data points for $0.9~\GeV \leq  W \leq 2.4~\GeV$ are
the unfolded results where the bin widths $\Delta W$ are 0.02~GeV
(0.04~GeV) in $W$ above (below) 1.6~GeV.
For the data points above 2.4~GeV, we average five data 
points each with a bin width of 0.02~GeV and obtain results
for every 0.1~GeV bin in $W$. 
We have removed the bins in
the range 3.3~GeV $\leq W \leq  3.6$~GeV, because we cannot separate 
the contributions from $\chi_{c0}$, $\chi_{c2}$  
and the continuum in a model-independent way, due to the finite 
mass resolution and insufficient statistics in this $W$ range.

\subsection{Systematic Errors in the Cross Sections}
We summarize the evaluation of the systematic errors
for $\sigma(|\cos \theta^*| \leq 0.8)$ ($\sigma(|\cos \theta^*| \leq 0.6)$ 
for $W \leq 0.72$~GeV) at each $W$ point. 
These come from the following sources: trigger efficiency,
reconstruction efficiency, $p_t$-balance cut, background subtraction,
beam-background effect, other efficiency errors including smoothing procedure,
unfolding procedure and luminosity function.

%\noindent
{\it Trigger efficiency}:  
the systematic error from
the Clst4 trigger is assigned as $2/3$ of the difference
of the efficiencies with different threshold
assumptions for the ECL cluster --
110~MeV and 100~MeV -- set in the trigger simulator for the
energy region $W \leq 2.5$~GeV.
We include a separate 4\% uncertainty in the HiE trigger
efficiency for the whole $W$ region. The systematic
errors from the two triggers are added in quadrature.
This systematic error becomes large in the low $W$ region, 20\%-30\% 
for $W \leq 0.8$~GeV.

%\noindent
{\it Reconstruction efficiency}:
the uncertainty in the $\pi^0$ reconstruction efficiency is estimated 
from a comparison of $D^0$-meson decays to $K^- \pi^+$ and
 $K^- \pi^+ \pi^0$.
An error of 6\% for two pions is assigned.

%\noindent
{\it The  $p_t$-balance cut} is 3\% - 5\%, which is 
one half of the correction discussed above.

%\noindent
{\it Background subtraction}:
 20\% of the size of the 
subtracted component is assigned as the error from this source.
In the $W$ region where the background subtraction 
is not applied ($W>1.2$~GeV), we neglect the error for 
1.2~GeV$ \leq W \leq 1.5$~GeV,
and assign 3\% for $W \geq 1.5$~GeV, which is an upper limit 
on the background contamination expected from the $p_t$-unbalanced
distributions.
For $W \geq 3.6$~GeV where we have applied a 3\% correction for
the background subtraction, we also assign a systematic error of
the same magnitude.

%\noindent
{\it Beam-background effect for event selection}:
we assign a 2\% - 4\% error depending on $W$ for 
uncertainties of the 
inefficiency in selection due to the effect of beam-background photons.

%\noindent
{\it Other efficiency errors:}
an extra error of 4\% is assigned for uncertainties in the efficiency
determinations based on the MC including the smoothing procedure.

%\noindent
{\it Unfolding procedure}:
we adopt the change in the unfolded yield in
each bin when we modify the effective-rank parameter 
(kset parameter) applied in the unfolding procedure~\cite{bib:svdunf}
in a reasonable range as a systematic error.

%\noindent
{\it Luminosity function}:
the uncertainty is estimated to be 4\% for $W < \; 3.0$~GeV
and 5\% for $W > \; 3.0$~GeV.

The total systematic error is 10\% in the wide energy region,
1.0~GeV$\leq W \leq 3.4$~GeV. 
The error, which is dominated by the background subtraction,
becomes much larger for lower $W$,
15\% at $W=0.85$~GeV, 30\% at $W=0.70$~GeV and 55\% at $W=0.61$~GeV.
For higher $W$, the systematic error is rather stable, remaining
at the 11\% level for 3.4~GeV$\leq W \leq 4.0$~GeV.

\subsection{Comparison of Cross Sections with the Previous Experiment}
\begin{figure}[ht]
\centering
{\epsfig{file=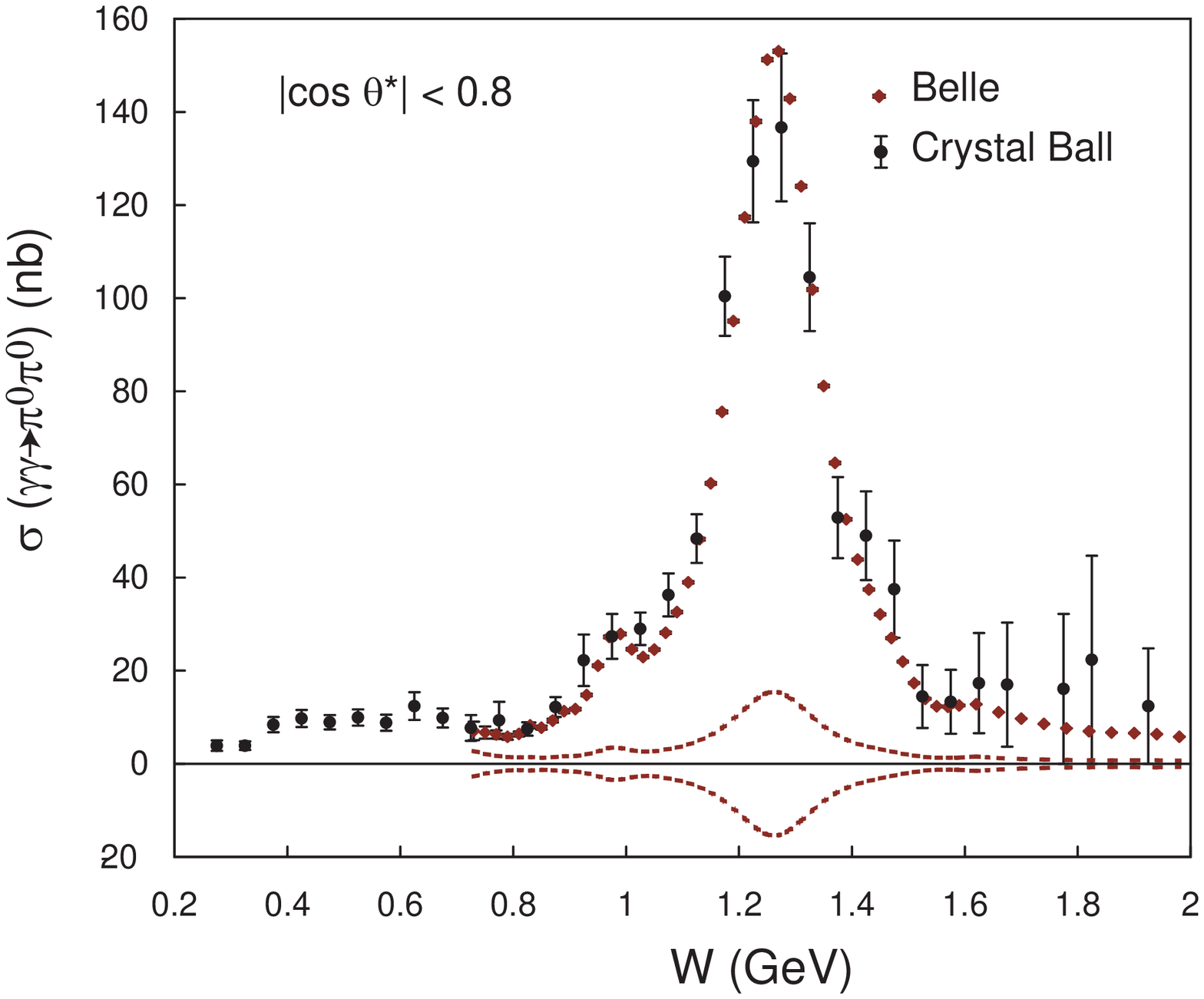, width=11cm}}
\centering
\caption{Cross section integrated over the angular region
$|\cos \theta^*| \leq 0.8$  
compared with the previous measurement from the 
Crystal Ball~\cite{bib:prev2}.
Dashed lines show the total systematic errors for the Belle measurement.
Systematic errors of similar size (not shown) are present in the Crystal Ball 
measurement.}
\label{fig:totcsb}
\end{figure}

The total cross section ($|\cos \theta^*| \leq 0.8$) is compared
with the previous measurement by
Crystal Ball at DORIS II~\cite{bib:prev2} (Fig.~\ref{fig:totcsb}). The 
agreement is fairly good.
The error bars shown are statistical only, and
the systematic errors (7\% for $W > 0.8~\GeV$ and
11\% for $W < 0.8~\GeV$ for the Crystal Ball results) 
should also be considered in the comparison. 
The present measurement has several hundred times more statistics than 
the Crystal Ball measurement.

\section{Fitting Differential Cross Sections}
\label{sec:fitting}
In this section, differential cross sections at each $W$ bin
are fitted to obtain information on the partial waves.
A simple parameterization is then fit to differential cross 
sections to obtain resonance parameters
of the $f_0(980)$ and another scalar meson denoted here as $f_0(Y)$
as well as to extract the $f_2(1270)$ fraction in the D$_0$ wave.

\subsection{Formalism}
In this channel, only partial waves of even angular momenta contribute.
Furthermore, in the energy region $W \simlt 3~\GeV$, $J > 4$ partial 
waves (next is $J=6$) may be neglected so that only S, D and G waves are 
to be considered.
The differential cross section can then be expressed as
\begin{equation}
\frac{d \sigma}{d \Omega} (\gamma \gamma \to \pi^0 \pi^0)
 = \left| S \: Y^0_0 + D_0 \: Y^0_2  + G_0 \: Y^0_4 \right|^2 
+ \left| D_2 \: Y^2_2  + G_2 \: Y^2_4 \right|^2 \; ,
\label{eqn:diff}
\end{equation}
where $D_0$ and $G_0$ ($D_2$ and $G_2$) denote the helicity 0 (2) components
of the D and G waves, respectively, and $Y^m_J$ are the spherical harmonics:
\begin{eqnarray}
Y^0_0 &=& \sqrt{\frac{1}{4 \pi}} \; , \nonumber \\
Y^0_2 &=& \sqrt{\frac{5}{16 \pi}}(3 \cos^2 \theta^* - 1) \; , \nonumber \\
\left| Y^2_2 \right| &=& \sqrt{\frac{15}{32 \pi}} \sin^2 \theta^*  \nonumber \\
Y^0_4 &=& \frac{3}{16} \sqrt{\frac{1}{\pi}} 
(35 \cos^4 \theta^* - 30 \cos^2 \theta^* +3) \; , \nonumber \\
\left| Y^2_4 \right| &=& \frac{3}{8} \sqrt{\frac{5}{2 \pi}} 
(7 \cos^2 \theta^* - 1) \sin^2 \theta^* \;.
\label{eqn:sphe}
\end{eqnarray}
Since the $|Y^m_J|$s are not independent,
%\begin{equation}
%|Y^2_2| = \frac{\sqrt{5} Y^0_0 - Y^0_2}{\sqrt{6}}, 
%\label{eqn:sphe2}
%\end{equation}
partial waves cannot be separated using measurements of
differential cross sections alone.

To overcome this problem,
we begin by rewriting Eq.~(\ref{eqn:diff}) as:
\begin{equation}
\frac{d \sigma}{4 \pi d |\cos \theta^*|} (\gamma \gamma \to \pi^0 \pi^0)
 = \hat{S}^2 \: |Y^0_0|^2  + \hat{D}_0^2 \: |Y^0_2|^2
+ \hat{D}_2^2  \: |Y^2_2|^2 \, 
+ \hat{G}_0^2  \: |Y^0_4|^2  \, 
+ \hat{G}_2^2  \: |Y^2_4|^2  \, .
\label{eqn:diff2}
\end{equation}
The amplitudes $\hat{S}^2$, etc. correspond to the cases where interference 
terms are neglected.
When interference terms are included, they can be written as (see
Appendix~\ref{sec:appendix}):
\begin{eqnarray}
\hat{S}^2 &=& |S|^2 + \sqrt{5} \Re{(S^* D_0)} 
- 4 \Re{(S^* G_0)}   + \frac{7}{\sqrt{5}} \Re{(D_0^* G_0)}
+ \frac{14 \sqrt{3}}{5} \Re{(D_2^* G_2)} 
\; , \nonumber \\
\hat{D}_0^2 &=& |D_0|^2 + \frac{1}{\sqrt{5}} \Re{(S^* D_0)} 
+ 2 \Re{(S^* G_0)}   + \frac{1}{\sqrt{5}} \Re{(D_0^* G_0)}
- \frac{4 \sqrt{3}}{5} \Re{(D_2^* G_2)} 
\; , \nonumber \\
\hat{D}_2^2 &=& |D_2|^2 - \frac{6}{\sqrt{5}} \Re{(S^* D_0)}
+ 2 \Re{(S^* G_0)}   + \sqrt{5} \Re{(D_0^* G_0)}
- \frac{9 \sqrt{3}}{5} \Re{(D_2^* G_2)} \; , \nonumber \\
\hat{G}_0^2 &=& |G_0|^2 + \frac{3}{\sqrt{5}} \Re{(D_0^* G_0)}
+ \frac{\sqrt{3}}{5} \Re{(D_2^* G_2)} \; , \nonumber \\
\hat{G}_2^2 &=& |G_2|^2 - \sqrt{5} \Re{(D_0^* G_0)}
- \frac{1}{\sqrt{3}} \Re{(D_2^* G_2)} \; .
\label{eqn:def1}
\end{eqnarray}
Since squares of spherical harmonics are independent of each other,
we can fit differential cross sections at each $W$ to obtain 
$\hat{S}^2$, $\hat{D}_0^2$, $\hat{D}_2^2$, $\hat{G}_0^2$
and $\hat{G}_2^2$.
The fit up to $J = 4$ is called the ``$SDG$ fit''.
At low energy, we expect that $J=4$ waves are unimportant.
Therefore we also perform a separate fit setting $\hat{G}_0^2 = \hat{G}_2^2 =0$,
which is called the ``$SD$ fit''. 

The unfolded differential cross sections are fitted, 
where only statistical errors are taken into account in the fit.
Although they are not independent at each $W$ because of the unfolding 
procedure, we treat them as independent in the fit.
The effect of correlations between bins is taken into account in systematic 
errors as described below.
Differential cross sections for $|\cos \theta^*| \leq 0.8$ are available for
$0.73~\GeV \leq W \leq 4.0~\GeV$.

\begin{figure}[ht]
 \centering
   {\epsfig{file=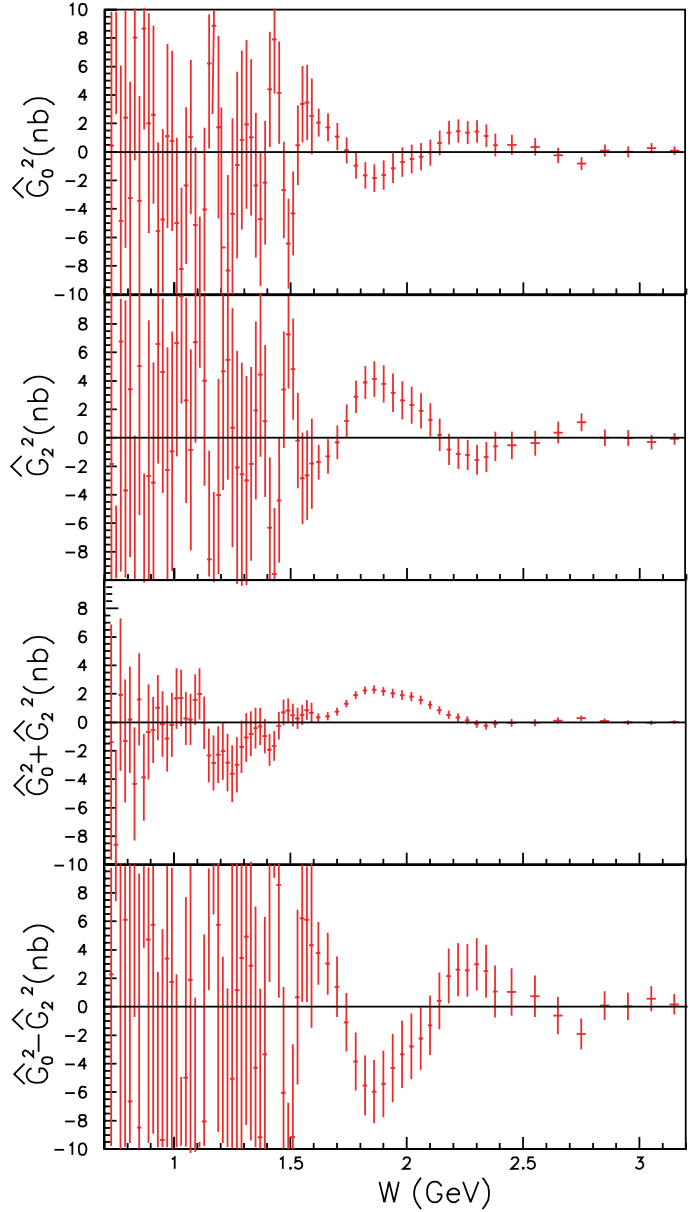,width=90mm}}
 \caption{The $\hat{G}_0^2$, $\hat{G}_2^2$
and $\hat{G}_0^2 \pm \hat{G}_2^2$ spectra for the $SDG$ fit.}
\label{fig:fit5gall}
\end{figure}

\begin{figure}[ht]
 \centering
   {\epsfig{file=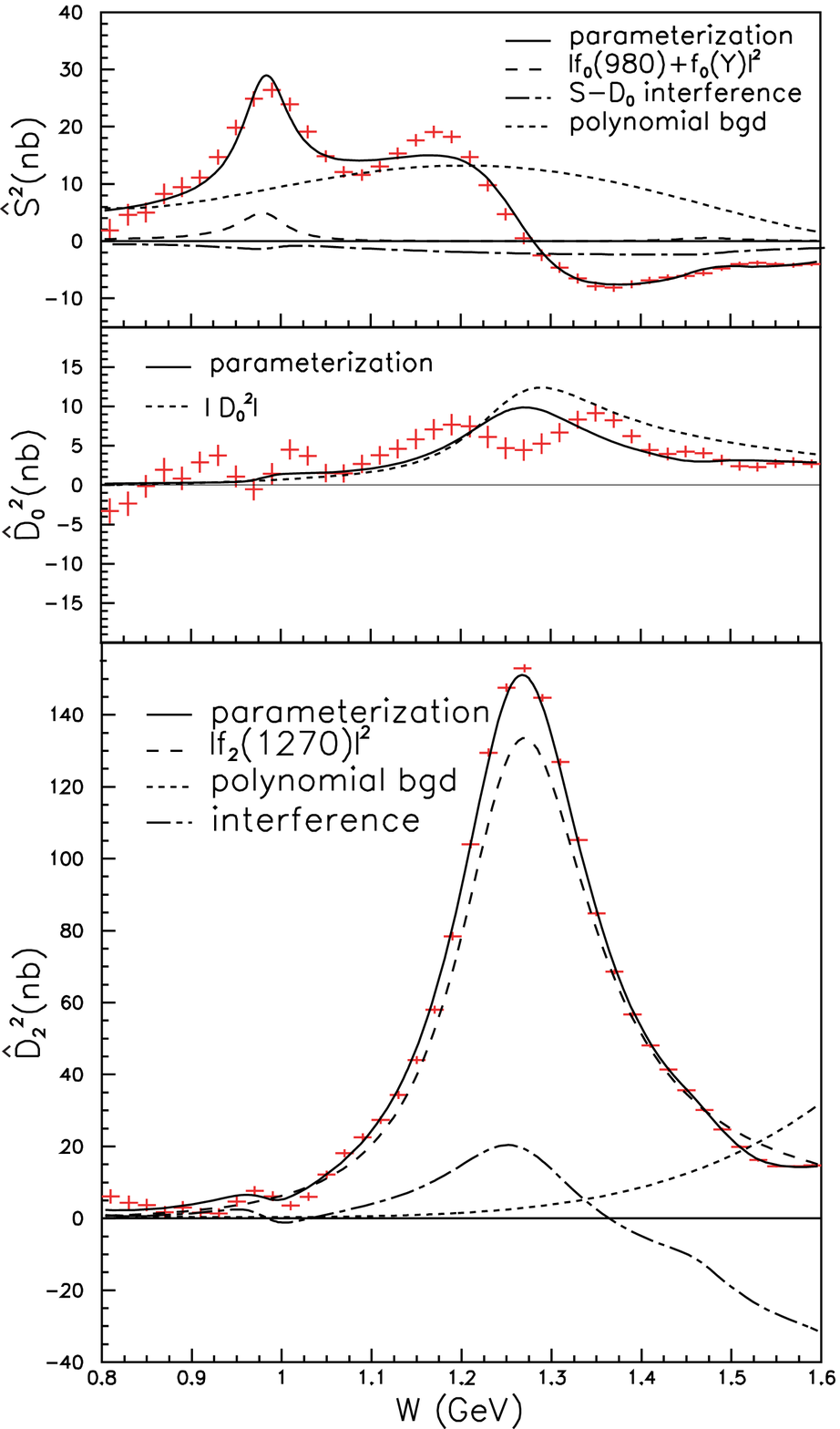,width=90mm}}
 \caption{The $\hat{S}^2$ (top), $\hat{D}_0^2$ (middle) and 
$\hat{D}_2^2$ (bottom) spectra.
Curves are result of the fit with the parameterization discussed in text.
The error bars are diagonal statistical 
errors only.}
\label{fig:a3nall}
\end{figure}

Examples of the fit quality are shown in Fig.~\ref{fig:dssp}.
In the energy region near $W=2~\GeV$, a need for $J=4$ waves is evident.
There, $\hat{G}_2^2$ deviates from zero as
can be seen in Fig.~\ref{fig:fit5gall}.
Since the behaviors of $|Y_4^0|^2$ and $|Y_4^2|^2$ are rather similar for
$|\cos \theta^*| \simlt 0.7$, we also fit with $|Y_4^0|^2 \pm |Y_4^2|^2$.
The bump in $\hat{G}_0^2 + \hat{G}_2^2$ may indicate the
presence of the $f_4(2050)$.
However, in the high energy region $W > 1.6$~GeV, there are many more 
resonances contributing and thus the model uncertainty becomes much larger.
Therefore in this paper we focus on the energy region 
$0.8~\GeV \leq W \leq 1.6~\GeV$, where G waves can be neglected.
The $\hat{S}^2$, $\hat{D}_0^2$ and $\hat{D}_2^2$ spectra are
shown in Fig.~\ref{fig:a3nall}.

\subsection{Fitting Partial Wave Amplitudes}
\label{sec:part}
%%Although the derived amplitudes $\hat{S}^2$, etc. are complicated functions of
%%partial waves $S$, etc. (Eq.~(\ref{eqn:def1})), they do give some indication
%%of their behavior.
%%Notably, the peak in $\hat{S}^2$ around $W=1.2~\GeV$ 
%%(Fig.~\ref{fig:a3nall}) would imply a contribution of a scalar
%%meson in the S wave, which could be the $f_0(1370)$
%%and/or a contribution from the $f_2(1270)$ that is 
%%also present in the D$_0$ wave.

In this section, we derive some information on the 
relevant resonances by fitting
differential cross sections by assuming certain functional forms for the
partial wave amplitudes to understand the general behavior of 
partial wave amplitudes and to check the consistency with the 
$\pi^+ \pi^-$ data~\cite{bib:mori1,bib:mori2}.
Note that we do not fit the obtained $\hat{S}^2$, etc. 
but fit the differential cross sections directly;
once the functional forms of amplitudes are assumed, we can use 
Eq.~(\ref{eqn:diff}) to fit differential cross sections. 

Here, we concentrate on the energy region $W \leq 1.6~\GeV$,
where  G$_0$ and G$_2$ can be neglected.
A full amplitude analysis will be performed in the near future using 
all data available in addition to the $\pi^0 \pi^0$ data. 
Thus we employ a simple parameterization in this study.
The $\hat{D}_2^2$ spectrum is dominated by the $f_2(1270)$ resonance.
%According to Clebsch-Gordan coefficient calculation,
%the ratio $D_0/D_2 = 1/6$ if we assume that the $\gamma \gamma$ system
%is in the S state.
The $f_2(1270)$ could  contribute to the D$_0$ wave.
The $\hat{S}^2$ distribution has peaks apparently due to the $f_0(980)$ 
resonance
and another resonance-like structure around 1.2~GeV, which motivates us
to introduce a scalar meson denoted here as $f_0(Y)$
together with a contribution from the $f_2(1270)$ in the D$_0$ wave.
The $f_0(Y)$ can be either the $f_0(1370)$ or $f_0(1500)$ or a
mixture of both~\cite{bib:PDG}.
Note that the mass and width of the $f_0(1370)$ 
are known with a large uncertainty~\cite{bib:PDG}.
Neither of these states has been observed previously in two-photon 
production~\cite{bib:aleph}.
The goal of this analysis is to obtain parameters of the $f_0(980)$ and
$f_0(Y)$, to check the consistency of the $f_2(1270)$ parameterization
where its two-photon branching fraction is floated
and to measure the helicity 0-to-helicity 2 ratio of $f_2(1270)$
production.

\subsubsection{Parameterization of Amplitudes}
Based on the above observations, the
S, D$_0$ and D$_2$ waves are parameterized as follows:
\begin{eqnarray}
S &=& A_{f_0(980)} e^{i \phi_{s0}} +  A_{f_0(Y)} e^{i \phi_{s1}} 
 + B_S , \nonumber \\
D_0 &=& \sqrt{\frac{r_{02}}{1+r_{02}}} A_{f_2(1270)} e^{i \phi_{d0}} 
+ B_{D0}, \nonumber \\
D_2 &=& \sqrt{\frac{1}{1+r_{02}}} A_{f_2(1270)} e^{i \phi_{d2}} 
+ A_{f_2'(1525)} e^{i \phi_{d2'}} + B_{D2},
\label{eqn:param}
\end{eqnarray}
where $A_{f_0(980)}$, $A_{f_0(Y)}$, $A_{f_2(1270)}$  and 
$A_{f_2'(1525)}$ are the amplitudes of 
the $f_0(980)$, another scalar resonance denoted as $f_0(Y)$, 
the $f_2(1270)$ and the $f_2'(1525)$, respectively;
$B_S$, $B_{D0}$ and $B_{D2}$ are ``background'' amplitudes for 
S, D$_0$ and D$_2$ waves; and
$\phi_{s0}$, $\phi_{s1}$, $\phi_{d0}$, $\phi_{d2}$ and $\phi_{d2'}$
are the phases of resonances relative to background amplitudes.
The parameter $r_{02}$ represents the fraction of the $f_2(1270)$
component in the D$_0$ wave.
Here, as a default, we assume the presence of both the $f_0(Y)$ in the 
S wave and the $f_2(1270)$ in the D$_0$ wave.
We also study the cases where either $r_{02}=0$ or there is no $f_0(Y)$.

We assume background amplitudes to be quadratic in $W$ for all the waves:
\begin{eqnarray}
B_S &=&  a_{sr} W^2  + b_{sr} W + c_{sr}
 + i (a_{si} W^2  + b_{si} W  + c_{si}) , \nonumber \\
B_{D0} &=& a_0 W^2 + b_0 W + c_0 , \nonumber \\
B_{D2} &=& a_2 W^2 + b_2 W + c_2 .
\label{eqn:para2}
\end{eqnarray}
The $D_0$ and $D_2$ background amplitudes
are taken to be real by definition.

We use the parameterization of the $f_0(980)$, $f_2(1270)$ and $f_2'(1525)$ 
given in Ref.~\cite{bib:mori1} and \cite{bib:mori2}.
We note that ${\B}(f_J \rightarrow \pi^0 \pi^0)/
{\B}(f_J \rightarrow \pi^+ \pi^-) = 1/2$ (because the $f_J$ mesons
are isoscalars).
For completeness, we reproduce the parameterization of the $f_0(980)$ and
$f_2(1270)$.
For the $f_0(980)$ meson, we take
\begin{equation}
A_{f_0(980)} = \frac{\sqrt{8 \pi \beta_{\pi}}}{W}
\frac{g_{f_0 \gamma\gamma}g_{f_0 \pi\pi}}{16\pi}
\cdot\frac{1}{D_{f_0}} ,
 \label{eqn:sigma}
\end{equation}
where $\beta_X = \sqrt{1-\frac{4 {M_X}^2}{W^2}}$ is the velocity of the 
particle $X$ with  mass $M_X$ in the two-body final states,
$g_{f_0XX}$ is related to the partial width
of the $f_0(980)$ meson via
$\Gamma_{XX} (f_0) = \frac{\beta_X g_{f_0 XX}^2}{16 \pi M_{f_0}}$.
The factor $D_{f_0}$ is given as follows~\cite{bib:denom}:
\begin{equation}
 D_{f_0}(W) = M_{f_0}^2 - W^2
              + \Re{\Pi_{\pi}^{f_0}}\lr{M_{f_0}}-\Pi_{\pi}^{f_0}\lr{W}
	      + \Re{\Pi_K^{f_0}}\lr{M_{f_0}} - \Pi_K^{f_0}\lr{W} ,
\nonumber
\end{equation}
where for $X = \pi$ or $K$,
\begin{equation}
 \Pi_X^{f_0}(W) =  \frac{\beta_X {g^2_{f_0XX}}}{16\pi}
       \left[i + \frac{1}{\pi}
			          \ln\frac{1-\beta_X}
				          {1+\beta_X}\right] .
\end{equation}
The factor $\beta_K$ is real in the region $W \geq 2M_K$
and becomes imaginary for $W < 2M_K$.
The mass difference between $K^{\pm}$ and $K^0$ $(\overline{K^0})$ is 
included by using $\beta_K = \frac{1}{2} (\beta_{K^{\pm}} + \beta_{K^0})$.
The parameters assumed and determined in Ref.~\cite{bib:mori1} are
summarized in Table~\ref{tab:f0fit}.
\begin{center}
\begin{table}[ht]
\caption{Parameters of the $f_0\lr{980}$ assumed and fitted in 
Ref.~\cite{bib:mori1}.}
\label{tab:f0fit}
\begin{tabular}{cccc} \hline \hline
Parameter & Value & Unit & Reference \\ \hline
 Mass & $985.6 ~^{+1.2}_{-1.5}\lr{\rm stat}
                   ~^{+1.1}_{-1.6}\lr{\rm syst}$ & $\MeV/c^2$
& \cite{bib:mori1}\\
$g_{f_0(980) \pi\pi}$ & $1.33 ~_{-0.23}^{+0.27}\lr{\rm stat}
                     ~_{-0.05}^{+0.16}\lr{\rm syst}$ & GeV 
& \cite{bib:mori1}\\
$\Gamma_{\pi^+\pi^-}$ & $34.2~^{+13.9}_{-11.8}\lr{\rm stat}
~^{+8.8}_{-2.5}\lr{\rm syst}$ & MeV & \cite{bib:mori1} \\
$g^2_{f_0(980) KK}/ g^2_{f_0(980) \pi\pi}$  
& $4.21 ~\pm 0.25 \lr{\rm stat}
                     ~\pm 0.21\lr{\rm syst}$ & -- & \cite{bib:bes}\\
$\Gamma_{\gamma\gamma}$ & $205
~^{+95}_{-83}\lr{\rm stat} ~^{+147}_{-117}\lr{\rm syst}$ & eV
& \cite{bib:mori1} \\
\hline \hline
\end{tabular}
\end{table}
\end{center}

Next, we give the parameterizations of the $f_2(1270)$ and $f_2'(1525)$ 
mesons.
The relativistic Breit-Wigner resonance amplitude
$A_R(W)$ for a spin-$J$ resonance $R$ of mass $m_R$ is given by
\begin{eqnarray}
A_R^J(W) &=& \sqrt{\frac{8 \pi (2J+1) m_R}{W}} 
%\nonumber \\
%&& 
\times \frac{\sqrt{ \Gamma_{\gamma \gamma}(W) \Gamma_{\pi^0 \pi^0}(W)}}
%{\cal B}(R \rightarrow \gamma \gamma) {\cal B}(R \rightarrow \pi^0 \pi^0)}} 
{m_R^2 - W^2 - i m_R \Gamma_{\rm tot}(W)} \; ,
\label{eqn:arj}
\end{eqnarray}
Hereafter we consider the case $J=2$ (the $f_2(1270)$ and $f_2'(1525)$ 
mesons).
The energy-dependent total width $\Gamma_{\rm tot}(W)$ is given by
\begin{equation}
\Gamma_{\rm tot}(W) = \sum_X \Gamma_{X \bar{X}} (W) \; ,
\label{eqn:gamma}
\end{equation}
where $X$ is $\pi$, $K$, $\gamma$, etc.
The partial width $\Gamma_{X \bar{X}}(W)$ is 
parameterized as~\cite{bib:blat}
\begin{equation}
\Gamma_{X \bar{X}} (W) = \Gamma_R {\cal B}(R \rightarrow X \bar{X}) 
\left( \frac{q_X(W^2)}{q_X(m_R^2)} \right)^5
\frac{D_2\left( q_X(W^2) r_R \right)}{D_2 \left( q_X(m_R^2) r_R \right)} \;,
\label{eqn:gamx}
\end{equation}
where $\Gamma_R$ is the total width at the resonance mass,
$q_X(W^2) = \sqrt{W^2/4 - m_X^2}$, $D_2(x) = 1/(9 + 3 x^2 +x^4)$,
and $r_R$ is an effective interaction radius that varies from 
1~$(\GeV/c)^{-1}$ to 7~$(\GeV/c)^{-1}$ in different hadronic 
reactions~\cite{bib:grayer}.
%For $X = \pi, \; K, \;{\rm and} \; \gamma$,
%the branching fractions are $0.848^{+0.025}_{-0.013}$,
%$0.046 \pm 0.004$, and $(1.41 \pm 0.13) \times 10^{-5}$, 
%respectively~\cite{bib:PDG}.
For the $4 \pi$ and the other decay modes,
$\Gamma_{4 \pi} (W) = \Gamma_R {\cal B}(R \rightarrow 4 \pi)
\frac{W^2}{m_R^2}$ is used instead of Eq.~(\ref{eqn:gamx}).
In  Ref.~\cite{bib:mori2}, all parameters of the $f_2(1270)$ are fixed at 
the PDG values~\cite{bib:PDG}, except for $r_R$, 
as summarized in Table~\ref{tab:f2fit}.

In the fit below, we float the branching fraction of the $f_2(1270)$
into two photons, because its value determined in the past experiments
is based on various assumptions.

\begin{center}
\begin{table}[ht]
\caption{Parameters of the $f_2\lr{1270}$ and $f_2'\lr{1525}$
 assumed}
\label{tab:f2fit}
\begin{tabular}{ccccc} \hline \hline
Parameter & $f_2(1270)$ & $f_2'(1525)$ & Unit & Reference \\ \hline
 Mass & $1275.4 \pm 1.1$ & $1525 \pm 5$ 
& $\MeV/c^2$ & \cite{bib:PDG}\\
$\Gamma_{\rm tot}$ & $185.2 ^{+3.1}_{-2.5}$ & $73^{+6}_{-5}$ 
& MeV & \cite{bib:PDG}\\
${\cal B}({f_2 \rightarrow \pi \pi})$ & $84.8^{+2.5}_{-1.3}$ 
& $0.82 \pm 0.15$ & \% &
\cite{bib:PDG} \\
${\cal B}({f_2 \rightarrow K \bar{K}})$ & $ 4.6 \pm 0.4$ 
& $88.8 \pm 3.1 $ & \% &
\cite{bib:PDG} \\
${\cal B}({f_2 \rightarrow \eta \eta})$ & -- 
& $ 10.3 \pm 3.1 $ & \% &
\cite{bib:PDG} \\
${\cal B}({f_2 \rightarrow \gamma \gamma})$ & $14.1 \pm 1.3$
& $1.11 \pm 0.14 $ & $\times 10^{-6} $ & \cite{bib:PDG}\\
$r_R$  & \multicolumn{2}{c}{$3.62 \pm 0.03$} & $(\GeV/c)^{-1}$
& \cite{bib:mori2} \\
\hline \hline
\end{tabular}
\end{table}
\end{center}

\begin{figure}
 \centering
   {\epsfig{file=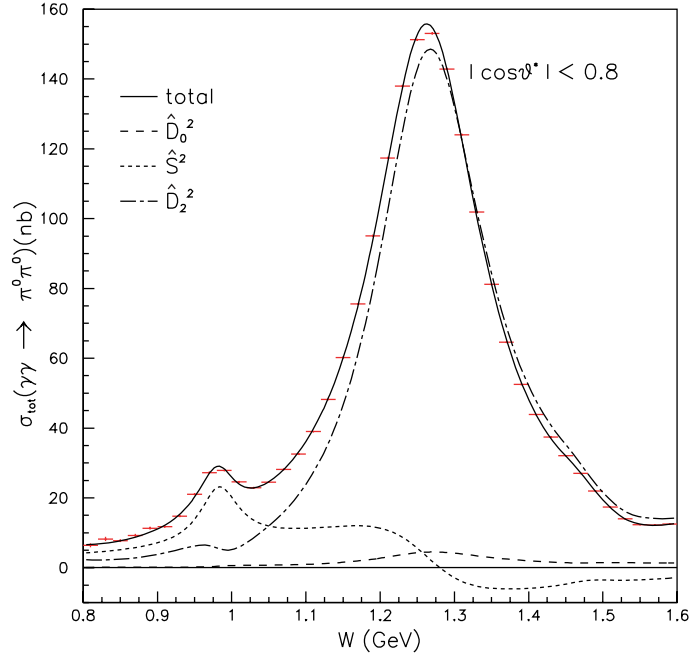,width=90mm}}
 \caption{Total cross section ($|\cos \theta^*| \leq 0.8$) and 
results of parameterization.
Contributions of $\hat{S}^2, \; \hat{D}_0^2$ and $\hat{D}_2^2$
are also shown.}
\label{fig:a3nt}
\end{figure}

Finally, the parameterization of the $f_0(Y)$ meson 
%(denoted $f_{0Y}$ here)
is taken to be:
\begin{equation}
A_{f_0(Y)} = \sqrt{\frac{8 \pi m_{f_0(Y)}} {W}}
\frac{ \sqrt{ \Gamma_{f_0(Y)} 
\Gamma_{\gamma \gamma} (f_0(Y)) {\cal B}(f_0(Y) \to \pi^0 \pi^0)}} 
{m^2_{f_0(Y)} -W^2 -i m_{f_0(Y)} \Gamma_{f_0(Y)}}
 \; ,
\label{eqn:f0y}
\end{equation}
where $\Gamma_{\gamma \gamma} (f_0(Y))$
is the two-photon width of the $f_0(Y)$ meson.

\subsubsection{Fitted Parameters}
We fit differential cross sections with the
parameterized amplitudes in the c.m. energy range
0.8~GeV$ \leq W \leq $ 1.6~GeV.
The parameter $g^2_{KK}/g^2_{\pi\pi}$ for the $f_0(980)$ is fixed to zero
because this is  preferred by the fit with a large error.
There are 25 parameters to be fitted.
The coefficients $a_0$ and $a_2$ in the continuum parameterizations 
are chosen to be positive to fix sign ambiguities.
About 3000 sets of randomly generated initial parameters are prepared 
and fitted using MINUIT~\cite{bib:minuit}
to search for the true minimum and to find any multiple solutions.
Once solutions are found,
several tens of MINUIT iterations are needed for convergence;
with many parameters (25 here and $21 - 26$ for later analyses); 
the approach to the minimum is rather slow.
A unique solution is found with $\chi^2/ndf = 1010.1/615 = 1.64$
for the nominal fit, where $ndf$ is the number of degrees of freedom.
The fitted parameters are listed in Table~\ref{tab:fit}.
The errors quoted are statistical only.
The fit quality is adequate, $\chi^2/ndf = 1.64$,
and represents the trend of the squared amplitudes as shown in 
Fig.~\ref{fig:a3nall}.
The error bars in the figure are diagonal statistical 
errors only.
The quantity $\hat{D}_2^2$ is well reproduced except below 1.1~GeV.
The effect of the $f_2'(1525)$ is rather small and
 not visible in the figures.
Additional assumptions or a more complicated model are
 needed to better reproduce the structures visible in 
$\hat{D}_0^2$ for the range $1.1~\GeV \simlt W \simlt 1.4~\GeV$ 
(the $f_2(1270)$ region).
 
\begin{center}
\begin{table}
\caption{Fitted parameters} 
\label{tab:fit}
\begin{tabular}{lcccc} \hline \hline
Parameter  & Nominal  & $r_{02}=0$ & No $f_0(Y)$ & Unit\\
\hline
Mass(${f_0(980)})$ 
&  $982.2 \pm 1.0$ & $980.2 \pm 1.0 $ & $983.7^{+1.5}_{-1.0}$ & MeV/$c^2$\\
$\Gamma_{\gamma \gamma}(f_0(980))$ 
&  $285.5^{+17.2}_{-17.1}$ &  $297.0^{+14.2}_{-13.7}$ 
& $370.5^{+20.2}_{-18.7}$ & eV\\
$g_{f_0(980) \pi \pi}$ 
& $1.82 \pm 0.03$ & $1.79 \pm 0.03$ & $1.89 \pm 0.03$ & GeV\\ \hline
Mass($f_0(Y))$ 
&  $1469.7 \pm 4.7$ & $1466.8 \pm 0.6$ & -- & MeV/$c^2$\\
$\Gamma(f_0(Y))$ 
& $89.7^{+8.1}_{-6.6}$ & $422.4^{+18.4}_{-19.8}$ & -- & MeV \\
$\Gamma_{\gamma \gamma} {\cal B}(f_0(Y) \to \pi^0 \pi^0)$ 
&  $11.2^{+5.0}_{-4.0}$ &  $6780.2^{+626.5}_{-574.7}$ & 0 (fixed) & eV \\
 \hline
$r_{02}$  & $3.69^{+0.24}_{-0.29}$ & 0 (fixed) & $5.04^{+0.26}_{-0.24}$ 
& \% \\
${\cal B} (f_2(1270) \rightarrow \gamma \gamma)$ 
&  $1.57 \pm 0.01$ &  $1.62^{+0.02}_{-0.01}$ 
& $1.52^{+0.13}_{-0.31}$ & $\times 10^{-5}$ \\
\hline
$\chi^2 \; (ndf)$ 
& 1010.1 (615)  & 1206.1 (617)  & 1253.3 (619)  & \\
\hline\hline
\end{tabular}
\end{table}
\end{center}

The total cross section ($|\cos \theta^*| \leq 0.8$) can be obtained
by integrating Eq.~(\ref{eqn:diff2}) as
\begin{equation}
\sigma_{\rm tot} = 0.8 \hat{S}^2 +  0.45728 \hat{D}_0^2
+ 0.988288 \hat{D}_2^2 \;,
\end{equation}
where the numerical factors come from the integration of spherical harmonics
for $|\cos \theta^*| \leq 0.8$.
The measured total cross section is compared with the prediction
obtained from the fitted amplitudes as shown in Fig.~\ref{fig:a3nt}.
They are reproduced reasonably well.

So far we assumed a need for both the $f_0(Y)$ in the S wave 
and the $f_2(1270)$ in the D$_0$ wave ($r_{02} \neq 0$).
We also study cases where either one of them is absent.
One thousand sets of randomly generated initial values are prepared and
fitted to find the true minimum.
Unique solutions are found in each case.
The values obtained are also listed in Table~\ref{tab:fit}.
The $\chi^2/ndf$ clearly favors the presence of both components.

Comparing the case of no $f_0(Y)$ and no $f_2(1270)$ in the D$_0$ wave,
the fit prefers the latter: $\chi^2/ndf = 1206.1/617$ compared to 
$1253.3/619$.
Thus we conclude that the possibility of only the $f_2(1270)$ in the D$_0$ 
wave is disfavored compared to the case of only the $f_0(Y)$ 
in the S wave by more than 6 standard deviations,
which is calculated from the difference of the $\chi^2$ taking the difference
of $ndf$s into account.

\subsubsection{Study of Systematic Errors}
The following sources of systematic errors on the parameters are 
considered: dependence on the fitted region, normalization errors in the 
differential cross sections,
assumptions on the background amplitudes,
uncertainties from the unfolding procedure, uncertainties in the parameters
assumed for the $f_0(980)$ and uncertainties in the measurements of 
the $f_2(1270)$ and $f_2'(1525)$.

In each study, a fit is made allowing all the parameters to float and
the differences of the fitted parameters from the nominal values
are quoted as systematic errors.
One thousand sets of randomly generated input parameters are again prepared
for each study
and fitted to search for the true minimum and for possible multiple solutions.
Unique solutions are found in each case.

Two fitting regions are tried: a narrow one 
($ 0.84~\GeV \leq W \leq 1.56~\GeV$)
and a wide one ($ 0.76~\GeV \leq W \leq 1.64~\GeV$).
Normalization error studies are divided into those from uncertainties of 
the overall normalization and those from smearing of the spectra in either
$|\cos \theta^*|$ or $W$.
For overall normalization errors, fits are made with two sets of values
of differential cross sections
obtained by multiplying by $(1 \pm \sigma_{\epsilon(W, |\cos \theta^*|)})$, 
where $\sigma_{\epsilon}$ is the relative efficiency error.
For smearing studies, $\pm 4$\% errors are assigned and differential cross 
sections are smeared by
$(1 \pm 0.1 |\cos \theta^*| \mp 0.04)$
and  $(1 \pm 0.1 W \mp 0.12)$.

For studies of background (BG) amplitudes, one of the waves 
is changed to either a first- or a third-order polynomial.
In estimating uncertainties from the unfolding procedure,
the kset parameter~\cite{bib:svdunf} is varied by $\pm 4$ from 
its nominal value.
When this parameter is increased, the constraints between adjacent bins become
weaker and oscillating solutions tend to appear as
the statistical errors increase.
When kset decreases, the opposite trend occurs.

For the parameterization of the $f_0(980)$, the fit prefers zero
for the ratio $g^2_{K K}/g^2_{\pi \pi}$ as stated above and hence we
set the ratio to be zero in the nominal fit. 
A systematic error due to this assumption is studied by setting
the ratio to be 4.45, a value 1 standard deviation larger
according to the BES measurement~\cite{bib:bes}.
Finally, the $f_2(1270)$ and $f_2'(1525)$ parameters 
are varied by their errors.

The resulting systematic errors are summarized in Table~\ref{tab:syser}.
It is noted that the mass of the $f_0(Y)$ jumps from $\sim 1.5~\GeV/c^2$
to $\sim 1.2~\GeV/c^2$, which is ``compensated'' by an increased value of
$r_{02}$ in a systematic error study on unfolding.
There, the change of cross-section values from the nominal ones is 
rather small (at most 15\% at $|\cos \theta^*| \simeq 0.8$) with some 
systematic dependence on $\cos \theta^*$ in some $W$ bins, 
showing the sensitive nature of this kind of analysis.
Total systematic errors are calculated by adding individual errors in 
quadrature.
{\large
\begin{table}
\caption{Systematic uncertainties for $f_0$ and $f_2$ parameters}
\label{tab:syser}
\begin{center}
\begin{tabular}{l|lll|lll|ll} \hline \hline 
& \multicolumn{3}{c|}{$f_0(980)$} & \multicolumn{3}{c|}{$f_0(Y)$} 
& \\ \cline{2-7}
Source & Mass & $\Gamma_{\gamma \gamma}$ & $g_{\pi \pi}$
 & Mass & $\Gamma_{\rm tot} $& $\Gamma_{\gamma \gamma} {\cal B}_{\pi^0 \pi0}$
 & ~~$r_{02}$ & ${\cal B} (f_2 \rightarrow \gamma \gamma) $  \\
& (MeV/$c^2$) & ~~~(eV)~~~ &~~(GeV)~~&(MeV/$c^2$)&(MeV)&(eV) & ~~(\%)~~~ 
 & $(\times 10^{-5})$ \\ \hline 
$W$-range & $^{+0.0}_{-2.8}$ & $^{+118.8}_{-0.0}$ & $^{+0.10}_{-0.02}$ 
& $^{+30.0}_{-49.7}$ & $^{+1.4}_{-13.0}$ & $^{+7.5}_{-5.0}$  
& $^{+0.51}_{-0.00}$  & $^{+0.57}_{-0.00}$ \\
Normalization& $^{+0.6}_{-0.4}$ & $^{+0.0}_{-28.5}$ & $^{+0.00}_{-0.01}$ 
 & $^{+8.8}_{-0.0}$ & $^{+15.0}_{-3.2}$ & $^{+10.5}_{-1.5}$ 
& $^{+0.29}_{-1.61}$  & $^{+0.44}_{-0.14}$ \\ 
Bias:$|\cos \theta^*|$ & $^{+0.0}_{-1.0}$ & $^{+6.9}_{-28.9}$ 
& $^{+0.00}_{-0.02}$ 
 & $^{+16.0}_{-0.0}$ & $^{+13.1}_{-0.0}$ & $^{+8.5}_{-0.0}$ 
& $^{+0.00}_{-1.09}$  & $^{+0.38}_{-0.00}$ \\
Bias:$W$& $^{+0.1}_{-2.2}$ & $^{+10.8}_{-0.0}$ & $^{+0.00}_{-0.01}$ 
 & $^{+25.8}_{-2.5}$ & $^{+1.0}_{-2.4}$ & $^{+8.2}_{-0.3}$ 
& $^{+0.01}_{-0.15}$  & $^{+0.58}_{-0.01}$ \\  \hline 
BG:$\Re (S)$ & $^{+0.0}_{-1.3}$ & $^{+127.1}_{-9.3}$ & $\pm 0.02$ 
 & $^{+6.8}_{-19.5}$ & $^{+0.3}_{-3.1}$ & $^{+2.2}_{-3.8}$ 
& $^{+1.14}_{-1.66}$  & $^{+0.07}_{-0.01}$ \\
BG:$\Im (S)$ & $^{+0.0}_{-2.5}$ & $^{+35.3}_{-0.0}$ & $^{+0.00}_{-0.01}$ 
 & $^{+19.8}_{-0.0}$ & $^{+0.0}_{-7.6}$ & $^{+0.8}_{-0.6}$ 
& $^{+2.52}_{-0.00}$ & $^{+0.02}_{-0.01}$  \\
BG:$D_0$ & $^{+0.1}_{-2.6}$ & $^{+30.6}_{-5.3}$ & $^{+0.01}_{-0.00}$ 
 & $^{+27.1}_{-2.2}$ & $^{+1.3}_{-1.7}$ & $^{+8.4}_{-0.5}$ 
& $^{+0.26}_{-0.62}$ & $^{+0.59}_{-0.00}$  \\
BG:$D_2$ & $^{+0.0}_{-2.9}$ & $^{+30.7}_{-0.0}$ & $^{+0.01}_{-0.00}$ 
 & $^{+42.6}_{-0.0}$ & $^{+6.3}_{-10.6}$ & $^{+10.0}_{-0.0}$ 
& $^{+0.00}_{-0.38}$ & $^{+0.51}_{-0.00}$  \\ \hline 
Unfolding & $^{+0.0}_{-5.0}$ & $^{+2.4}_{-48.5}$ & $^{+0.22}_{-0.16}$ 
& $^{+0.0}_{-249.5}$ & $^{+39.6}_{-0.0}$ & $^{+602.9}_{-0.0}$ 
& $^{+15.63}_{-0.37}$ & $^{+0.33}_{-0.00}$ \\ 
$g_{KK} \neq 0$ & $^{+8.0}_{-0.0}$ & $^{+97.4}_{-0.0}$ & $^{+0.00}_{-0.02}$ 
 & $^{+10.5}_{-0.0}$ & $^{+12.7}_{-0.0}$ & $^{+11.1}_{-0.0}$  
& $^{+0.00}_{-0.91}$ & $^{+0.46}_{-0.00}$  \\ \hline
$f_2$:mass& $^{+0.1}_{-0.7}$ & $^{+26.5}_{-26.6}$ & $^{+0.02}_{-0.03}$  
 & $^{+5.1}_{-3.8}$ & $^{+12.3}_{-10.7}$ & $^{+5.7}_{-2.9}$  
& $^{+0.36}_{-0.21}$ &  $^{+0.00}_{-0.01}$  \\
$f_2$:width& $^{+0.6}_{-1.2}$ & $^{+3.9}_{-2.8}$ & $^{+0.00}_{-0.01}$ 
 & $^{+11.7}_{-10.3}$ & $^{+2.6}_{-0.5}$ & $^{+1.5}_{-0.0}$  
& $^{+0.13}_{-0.11}$ & $^{+0.01}_{-0.02}$\\ 
$f_2:\B(\pi \pi)$ & $^{+0.0}_{-0.1}$ & $^{+0.7}_{-0.0}$ & $ \pm 0.00$ 
 & $^{+0.4}_{-0.0}$ & $^{+1.7}_{-0.0}$ & $^{+0.7}_{-0.0}$  
& $^{+0.00}_{-0.07}$ & $^{+0.02}_{-0.00}$\\ 
$f_2:\B(K K)$ & $^{+0.0}_{-0.3}$ & $^{+4.0}_{-8.7}$ & $^{+0.00}_{-0.01}$ 
 & $^{+1.8}_{-0.0}$ & $^{+2.5}_{-0.0}$ & $^{+1.4}_{-0.1}$  
& $^{+0.02}_{-0.00}$ & $\pm 0.00$\\ 
$f_2:r_R$ & $ \pm 0.1 $ & $^{+26.5}_{-3.3}$ & $^{+0.02}_{-0.01}$ 
 & $^{+5.1}_{-0.0}$ & $^{+12.3}_{-0.0}$ & $^{+5.7}_{-0.0}$  
& $^{+0.36}_{-0.08}$ & $ \pm 0.00$\\ \hline
$f_2'$:mass& $^{+0.0}_{-0.1}$ & $^{+0.9}_{-0.5}$ & $^{+0.00}_{-0.00}$  
 & $^{+1.0}_{-0.6}$ & $^{+0.3}_{-0.0}$ & $^{+0.3}_{-0.1}$  
& $^{+0.05}_{-0.00}$ &  $ \pm 0.00 $  \\
$f_2'$:width& $ \pm 0.0 $ & $ \pm 0.5 $ & $ \pm 0.00 $ 
 & $^{+0.5}_{-0.3}$ & $^{+1.1}_{-0.4}$ & $^{+0.2}_{-0.0}$  
& $ \pm 0.00 $ & $ \pm 0.00 $\\ 
$f_2':\B(\pi \pi)$ & $ \pm 0.0 $ & $^{+1.3}_{-0.8}$ & $ \pm 0.00 $ 
 & $^{+2.3}_{-1.9}$ & $^{+0.2}_{-0.9}$ & $^{+0.9}_{-0.8}$  
& $^{+0.05}_{-0.02}$ & $ \pm 0.00 $\\ 
$f_2':\B(K K)$ & $ \pm 0.0 $ & $^{+0.5}_{-0.0}$ & $ \pm 0.00 $ 
 & $ \pm 0.1 $ & $^{+0.2}_{-0.0}$ & $^{+0.1}_{-0.0}$  
& $ \pm 0.00 $ & $ \pm 0.00 $\\ 
$f_2':\B(\gamma \gamma )$ & $ \pm 0.0 $ & $^{+0.7}_{-0.2}$ & $ \pm 0.00 $ 
 & $ \pm 1.4 $ & $^{+0.6}_{-0.7}$ & $^{+0.7}_{-0.6}$  
& $ \pm 0.01 $ & $ \pm 0.00 $\\ 
$f_2':r_R$ & $ \pm 0.0 $ & $^{+0.2}_{-0.1}$ & $ \pm 0.00 $ 
 & $ \pm 0.0 $ & $^{+0.1}_{-0.0}$ & $ \pm 0.0 $  
& $ \pm 0.00 $ & $ \pm 0.00 $\\ 
\hline
Total & $^{+8.1}_{-8.0}$ & $^{+210.9}_{-70.1}$ & $^{+0.24}_{-0.17}$ 
& $^{+72.1}_{-255.4}$ & $^{+49.9}_{-22.0}$  & $^{+603.4}_{-7.2}$ 
& $^{+15.89}_{-2.85}$ & $^{+1.39}_{-0.14}$\\
\hline  \hline 
\end{tabular}
\end{center}
\end{table}
}
\subsubsection{Summary of Fit Results}
Once the amplitudes are parameterized, differential cross sections can be
fitted to obtain the parameters as described above.
The results are much more powerful than a simple fit to
the total cross section.
This is because there are so many points available that provide
rich information.
Although the fit quality is not very good as can be seen from 
$\chi^2/ndf = 1.64$, the fit is stable despite the fact that the approach to
the minimum is slow requiring tens of MINUIT iterations.

In Tables~\ref{tab:fitf0} and \ref{tab:fits'}
the results obtained for $f_0(980)$ and $f_0(Y)$ are summarized 
and compared with the PDG~\cite{bib:PDG} and with previous 
measurements~\cite{bib:mori1, bib:cball}.
The $f_0(980)$ parameters obtained here are consistent
with those obtained from Belle's measurement of 
$\pi^+ \pi^-$~\cite{bib:mori1}.
The $f_0(Y)$ fitted mass is close to the $f_0(1500)$ mass, but
is also consistent with the $f_0(1370)$ mass because of the large systematic
error in this experiment and the large uncertainty in the $f_0(1370)$ mass
from the PDG~\cite{bib:PDG}.
Although the product $\Gamma_{\gamma \gamma} {\cal B}(\pi^0 \pi^0)$ is 
consistent with zero given the size of the systematic error, 
the possibility that
$\Gamma_{\gamma \gamma} {\cal B}(\pi^0 \pi^0) = 0$  is disfavored
according to the fit 
(by comparing the nominal fit and the fit with no $f_0(Y)$ in Table~III).

The branching fraction of the $f_2(1270)$ to two photons is measured to be
$(1.57 \pm 0.01~^{+1.39}_{-0.14}) \times 10^{-5}$ in good agreement with
the value $(1.41 \pm 0.13) \times 10^{-5}$ in the PDG~\cite{bib:PDG}.
The value of $r_{02}$, the helicity 0-to-helicity 2 ratio of the $f_2(1270)$,
is $r_{02} = (3.7~\pm 0.3 ~^{+15.9}_{-2.9})$\%.
This is the first measurement that does not neglect
interference.
However, the large systematic errors in the $f_0(Y)$ parameters and $r_{02}$
indicate the subtle nature of this kind of fitting.
We find that the parameters of the $f_0(Y)$ and the 
$f_2(1270)$ in D$_0$ shown in Table III are rather strongly correlated. There seem 
to be structures that require helicity=0 components both around 1.2 GeV 
and 1.4 GeV. The structure near 1.2 GeV is more prominent, which is supported 
by the relatively robust $r_{02}$ value.  When a different unfolding solution 
is fitted, the mass of the $f_0(Y)$ jumps from 1.4 GeV to 1.2 GeV due to small 
changes in the cross sections and their statistical errors. It is difficult to 
disentangle these two structures just based on a simple model. We simply quote 
large systematic errors for the $f_0(Y)$ parameters.

\begin{center}
\begin{table} [h]
\caption{Fitted parameters of the $f_0(980)$} 
\label{tab:fitf0}
\begin{tabular}{c|cccc} \hline \hline
Parameter & Belle($\pi^0 \pi^0$) & Belle($\pi^+ \pi^-$) & PDG & Unit\\ \hline
Mass  & $982.2 \pm 1.0~^{+8.1}_{-8.0} $
& $985.6~^{+1.2}_{-1.5}~^{+1.1}_{-1.6}$ & $980 \pm 10$ & $\MeV/c^2$\\
$\Gamma_{\gamma \gamma}$ 
& $286 \pm 17~^{+211}_{-70} $
& $205~^{+95}_{-83}~^{+147}_{-117} $
& $310~^{+80}_{-110}$ & eV \\
$\Gamma_{\pi \pi}$ & $66.9 \pm 2.2 ~^{+17.6}_{-12.5} $
& $51.3~^{+20.9}_{-17.7}~^{+13.2}_{-3.8} $
& Unknown & MeV \\ \hline \hline
\end{tabular}
\end{table}
\end{center}
\begin{center}
\begin{table}
\caption{Fitted parameters of the $f_0(Y)$} 
\label{tab:fits'}
\begin{tabular}{c|ccccc} \hline \hline
Parameter & Belle$(\pi^0 \pi^0)$
& Crystal Ball 
& $f_0(1370)$(PDG) & $f_0(1500)$(PDG) 
& Unit \\ \hline
Mass  & $1470~^{+6}_{-7}~^{+72}_{-255} $
& 1250 
& 1200 - 1500 & $1507 \pm 5$ & $\MeV/c^2$\\
$\Gamma_{\rm tot}$  & $90~^{+2}_{-1}~^{+50}_{-22} $
& $ 268 \pm 70 $ 
& 150 - 200 & $109 \pm 7$ & MeV \\
$\Gamma_{\gamma \gamma} {\cal B}(\pi^0 \pi^0)$ 
& $11~^{+4}_{-2}~^{+603}_{-7}$ 
& $430 \pm 80 $ 
& Unknown & Not seen & eV \\ \hline \hline
\end{tabular}
\end{table}
\end{center}

\section{Summary and Conclusion}
\label{sec:summary}
We present the total and differential 
cross sections for the process
$\gamma \gamma \rightarrow \pi^0 \pi^0$ for $0.6~\GeV \leq W \leq 4.0~\GeV$
with the Belle detector at the KEKB asymmetric-energy $e^+ e^-$ collider. 
The 95~fb$^{-1}$ data sample has several hundred 
times higher statistics
than the previous measurements.
The differential cross sections are measured up to $|\cos \theta^*| = 0.8$,
which gives high sensitivity to the behavior of amplitudes.
A clear peak corresponding to the $f_0(980)$ is observed besides the
dominant $f_2(1270)$ and a dip-peak structure around $W=1.6$~GeV in the
total cross section. A general behavior of amplitudes is studied by
fitting the differential cross sections in a simple model, which includes
the S, D$_0$ and D$_2$ waves that are parameterized as smooth
backgrounds and resonances:
%%In this paper, the differential cross sections are fitted in the region 
%%0.8~GeV$ \leq W \leq $ 1.6~GeV to obtain information on S, D$_0$
%%and D$_2$ waves in a simple model where the amplitudes consist of a smooth 
%%background and resonances: 
the $f_0(980)$, $f_2(1270)$, $f_2'(1525)$
and another possible scalar resonance.

We obtain a reasonable fit with the $f_2(1270)$ parameters fixed at
the world-average values and its two-photon width floating. The fit yields its value, 
which is consistent with the world-average one. The $f_0(980)$ parameters fitted are
consistent with the values determined in the $\pi^+ \pi^-$ analysis~\cite{bib:mori1,bib:mori2}. 
Note that the latter are obtained just by fitting the total cross section,
while in this paper we fit differential cross sections. The structure in
$\hat{S}^2$ around 1.2 GeV can be reproduced by the fraction of the
$f_2(1270)$ present in the D$_0$ wave and/or the $f_0(Y)$. However, we cannot 
disentangle them more clearly.

\section*{Acknowledgment}
We thank the KEKB group for the excellent operation of the
accelerator, the KEK cryogenics group for the efficient
operation of the solenoid, and the KEK computer group and
the National Institute of Informatics for valuable computing
and SINET3 network support. We acknowledge support from
the Ministry of Education, Culture, Sports, Science, and
Technology of Japan and the Japan Society for the Promotion
of Science; the Australian Research Council and the
Australian Department of Education, Science and Training;
the National Natural Science Foundation of China under
contract No.~10575109 and 10775142; the Department of
Science and Technology of India; 
the BK21 program of the Ministry of Education of Korea, 
the CHEP SRC program and Basic Research program 
(grant No.~R01-2005-000-10089-0) of the Korea Science and
Engineering Foundation, and the Pure Basic Research Group 
program of the Korea Research Foundation; 
the Polish State Committee for Scientific Research; 
%-> remove for now: under contract No.~2P03B 01324; 
the Ministry of Education and Science of the Russian
Federation and the Russian Federal Agency for Atomic Energy;
the Slovenian Research Agency;  the Swiss
National Science Foundation; the National Science Council
and the Ministry of Education of Taiwan; and the U.S.\
Department of Energy.

\appendix
\section{Interfering Amplitudes}
\label{sec:appendix}
In order to obtain Eq.~(\ref{eqn:def1}), we express
$Y^0_2$, $Y^0_4$, $Y^0_2 Y^0_4$ and  $|Y^2_2| |Y^2_4|$,
in terms of $|Y^0_0|^2$,  $|Y^0_2|^2$,  $|Y^2_2|^2$,  $|Y^0_4|^2$,
and $|Y^2_4|^2$.
We use a power series in $\cos^2 \theta^*$.
The functions, $|Y^0_0|^2$,  $|Y^0_2|^2$,  $|Y^2_2|^2$,  $|Y^0_4|^2$,
and $|Y^2_4|^2$ can be written as
\begin{equation}
\left( \begin{array}{c}
4 \pi |Y^0_0|^2  \\ 4 \pi |Y^0_2|^2 \\ 4 \pi |Y^2_2|^2 \\
4 \pi |Y^0_4|^2 \\ 4 \pi |Y^2_4|^2
\end{array} \right) = 
\left( \begin{array}{ccccc}
1 & 0  & 0  & 0  & 0 \\
\frac{5}{4} & - \frac{15}{2} & \frac{45}{4} & 0 & 0 \\
\frac{15}{8} & - \frac{15}{4} & \frac{15}{8} & 0 & 0 \\
\frac{81}{64} & - \frac{405}{16} & \frac{4995}{32} 
& - \frac{4725}{16} & \frac{110255}{64} \\
\frac{45}{32} & - \frac{45}{2} & \frac{1755}{16} 
& - \frac{315}{2} & \frac{2205}{32} \\
\end{array} \right) 
\left( \begin{array}{c}
1 \\ \cos^2 \theta^* \\ \cos^4 \theta^* \\ \cos^6 \theta^* \\
\cos^8 \theta^* \\
\end{array} \right)
\label{eqn:mat1} 
\end{equation}
Equation~(\ref{eqn:mat1}) is inverted to obtain
\begin{equation}
\left( \begin{array}{c}
1 \\ \cos^2 \theta^* \\ \cos^4 \theta^* \\ \cos^6 \theta^* \\
\cos^8 \theta^* \\
\end{array} \right) =
\left( \begin{array}{ccccc}
1 & 0  & 0  & 0  & 0 \\
\frac{2}{3} &  \frac{1}{15} & -\frac{2}{15} & 0 & 0 \\
\frac{1}{3} & \frac{2}{15} & - \frac{4}{15} & 0 & 0 \\
\frac{16}{75} & \frac{73}{525} & - \frac{34}{175} 
& \frac{16}{1575} & - \frac{8}{315} \\
\frac{9}{175} & \frac{156}{1225} & - \frac{184}{1225} 
& \frac{256}{99225} & - \frac{32}{735} \\
\end{array} \right) 
\left( \begin{array}{c}
4 \pi |Y^0_0|^2  \\ 4 \pi |Y^0_2|^2 \\ 4 \pi |Y^2_2|^2 \\
4 \pi |Y^0_4|^2 \\ 4 \pi |Y^2_4|^2
\end{array} \right)
\label{eqn:mat2} 
\end{equation}
Equation~(\ref{eqn:def1}) can then be derived.

\end{document}